\documentclass[12pt]{article}
\usepackage{sectsty}
\usepackage{geometry}
\usepackage{graphicx}
\usepackage{indentfirst} 
\usepackage[english]{babel}
\usepackage[utf8x]{inputenc}
\usepackage{amsmath}
\usepackage{autobreak}
\usepackage{booktabs}
\usepackage{longtable}
\usepackage{array}
\usepackage{multirow}
\usepackage{wrapfig}
\usepackage{float}
\usepackage{colortbl}
\usepackage{pdflscape}
\usepackage{tabu}
\usepackage{threeparttable}
\usepackage{threeparttablex}
\usepackage[normalem]{ulem}
\usepackage{makecell}
\usepackage{graphicx}
\usepackage{setspace}
\usepackage{cite}
\usepackage{authblk}
\usepackage{xspace}

\title{Confidence Intervals for Ratios of Proportions in Stratified Bilateral Correlated Data}

\author[a]{Wanqing Tian}
\author[b]{Chang-Xing Ma \thanks{CONTACT Chang-Xing Ma: cxma@buffalo.edu, Department of Biostatistics, University at Buffalo, 3435 Main St., Buffalo, NY 14214, USA}}
\affil[a,b]{Department of Biostatistics, University at Buffalo, Buffalo, New York, USA}

\sectionfont{\small}
\subsectionfont{\small}
\subsubsectionfont{\small}

\begin{document}
\maketitle

\textbf{Abstract} 

Confidence interval (CI) methods for stratified bilateral studies use intraclass correlation to avoid misleading results. In this article, we propose four CI methods (sample-size weighted global MLE-based Wald-type CI, complete MLE-based Wald-type CI, profile likelihood CI, and complete MLE-based score CI) to investigate CIs of proportion ratios to clinical trial design with stratified bilateral data under Dallal’s intraclass model. Monte Carlo simulations are performed, and the complete MLE-based score confidence interval (CS) method yields a robust outcome. Lastly, a real data example is conducted to illustrate the proposed four CIs.

\section{Introduction}

In random clinical trials, we often encounter bilateral data in patients who receive the treatment based on paired body parts or organs (such as eyes, ears, kidneys, etc.). Although different studies from different aspects constructed statistical significance tests for bilateral data, the confidence interval (CI) construction becomes a trend, which provides more information than statistical significance tests. Several studies developed CIs for proportion ratio in two independent binomial proportions, such as Gart and Nam and Koopman  \cite{ gart1988approximate, koopman1984confidence}. Still, none of those studies take the intraclass correlation into consideration when dealing with bilateral data.

Ignoring correlations in the correlated data leads to error, which has been found in Sainani's study \cite{sainani2010importance}. To address this problem, Rosner proposed that a conditional probability of response at one side of paired body parts or organs gives a response at the other body parts or organs is a positive constant R time the response rate  \cite{Rosner_1982}. Related to this, Donner assumed that all treatment groups share one intraclass correlation coefficient $\rho$  \cite{Donner1989rhoModel}. Following this stream of literature, several studies have proposed constructing CIs under these two models with considerable progress. Under Rosner's R model, Peng et al. developed asymptotic CIs of proportion ratios in correlated paired data \cite{Peng_2019}. Tang et al. proposed CIs construction of the difference between two correlated proportions with missing observations \cite{tang2016confidence}. Tang et al. presented a confidence interval approach for comparative studies involving binary outcomes in paired organs \cite{ tang2013confidence }. Yang et al. constructed the CIs for many-to-one comparisons of proportion differences in correlated paired data \cite{ yang2021simultaneous}. Wang and Shan considered the relative risk and the odds ratio of exact CIs \cite{wang2015exact}. Under Donnar's model, Li and Ma constructed asymptotic tests and CIs for the odds ratio of two proportions in bilateral correlated data\cite{li2022statistical}. Pei et al. presented the CIs for correlated proportion differences from paired data in a two-arm randomized clinical trial \cite{pei2012confidence}.

However, Dallal criticizes Rosner’s assumption and points out that “the constant R model will give a poor fit if the characteristic is almost certain to occur bilaterally with widely varying group-specific prevalence” \cite{Dallal_1988}. Dallal believes the conditional probability of response at one side of paired body parts or organs, giving a response at the other body parts or organs is a constant $\gamma$ \cite{Dallal_1988}. 

Under the stratified bilateral correlated data framework, several studies have developed various CIs with different intraclass models. Xue and Ma developed CIs for proportion ratios using Rosner's R model \cite{Xue_2019SMMR}. Zhuang et al. constructed CIs for proportion ratios with Donner's  $\rho$ model \cite{Zhuang_2018CI}. 
Shen et al. presented  CIs for risk difference with Donner's  $\rho$ model \cite{shen2018CI}. Tang et al. introduced CIs for risk differences in stratified paired designs under Rosner's R model \cite{tang2013simultaneous}. However, constructing CIs under Dallal's intraclass model in stratified bilateral correlated data has not been explored yet, especially the CIs for proportion ratio. The primary goal of this article is to investigate CIs of proportion ratios to clinical trial design with stratified bilateral data under Dallal’s intraclass model.

The remainder of this paper is organized as follows: section 2 presents the data structure and hypotheses, and section 3 introduces the maximum-likelihood estimation under each hypothesis. We propose four confidence interval construction methods for proportion ratio under Dallal’s model in section 4. Accordingly, we investigate the performance and robustness of four confidence interval construction methods by using simulation studies in section 5. Section 6 uses a real example of the OME study to illustrate our proposed methods. Conclusions and future works are in section 7.

\section{Data structure and Hypotheses}

\subsection{Notation}

Let $m_{l i j}$ be the number of patients in the $i^{t h}$ group of the $j^{t h}$ stratum with $l^{t h}$ responses, where $i$ = 1,2, $j = 1, \ldots,J$ , $l$ = 0, 1, 2. $m_{+i j}=m_{0 i j}+m_{1 i j}+m_{2 i j}$ represent the number of patients in the $i^{t h}$ group of the $j^{t h}$ stratum. The $p_{l i j}$ is corresponding to the probability of patients in  $i^{t h}$ group of the $j^{t h}$ stratum with $l^{t h}$ responses. Define $Z_{i j k h}$ as the indicator of the response of $k^{t h}$ eyes of the $h^{t h}$ patient in the $i^{t h}$ group from $j^{t h}$ stratum, where $k$ = 1, 2 and $h=1,2, \ldots, m_{+i j}$. If $Z_{i j k h} = 1$ then the improvement response occurs, and  $Z_{i j k h} = 0$ otherwise. Therefore, donate $\operatorname{Pr}\left(Z_{h i j k}=1\right)= \pi_{i j},\left(0 \leq \pi_{i j} \leq 1\right)$ as probability of having response on one site.

$$
  \begin{array}{cccc}
\hline & \multicolumn{2}{c}{\text { Group }} \\
\cline { 2 - 3 } \text { Number of Responses }(l) & 1 & 2 & \quad \text { Total } \\
\hline 0 & m_{01 j}\left(p_{01 j}\right) & m_{02 j}\left(p_{02 j}\right) & m_{0+j} \\
1 & m_{11 j}\left(p_{11 j}\right) & m_{12 j}\left(p_{12 j}\right) & m_{1+j} \\
2 & m_{21 j}\left(p_{21 j}\right) & m_{22 j}\left(p_{22 j}\right) & m_{2+j} \\
\hline \text { Total } & m_{+1 j} & m_{+2 j} & m_{++j} \\
\hline
\end{array}
$$
  
We explore the intraclass correlation based on Dallal's model that is conditional probability of response at one side of paired body parts or organs, giving a response at the other body parts or organs is a constant $\gamma$. Therefore, we assume $P\left(Z_{h i j k}=1 \mid Z_{h i j(3-k)}=1\right)=\gamma_{j}$, and the probabilities of eye with none, one or both can be express as following:

$$p_{0 i j }=1-\left(2-\gamma_{  j}\right) \pi_{i j },$$

$$\quad p_{1 i j }=2 \pi_{i j }\left(1-\gamma_{ j}\right),$$

$$\quad p_{2 i j }=\pi_{i j} \gamma_{ j},$$
and $p_{0 i j}+p_{1 i j}+p_{2 i j}=1$ for any fixed $i$ and $j$. The joint likelihood function for the observed data $\boldsymbol{m_{i j}}=\left(m_{0 i j}, m_{1 i j}, m_{2 i j}\right)^{T}$ is given by 

$$f\left(\boldsymbol{m_{i j}} \mid \boldsymbol{p_{i j}}\right)=\prod_{j=1}^{J} \prod_{i=1}^{2} \frac{m_{+i j} !}{m_{0 i j} ! m_{1 i j} ! m_{2 i j} !} p_{0 i j}^{m_{0 i j}} p_{1 i j}^{m_{1 i j}} p_{2 i j}^{m_{2 i j}} , $$ where $\boldsymbol{p_{i j}}=\left(p_{0 i j}, p_{1 i j}, p_{2 i j}\right)^{T}$. The corresponding log-likelihood function can be expressed as:
$$
l\left( \boldsymbol{\pi_{i}}, \boldsymbol{\gamma}\right)=\sum_{j=1}^{J} \sum_{i=1}^{2}\left\{m_{0 i j} \log \left[1-\left(2-\gamma_{j}\right) \pi_{i j}\right]+m_{1 i j} \log \left[2 \pi_{i j}\left(1-\gamma_{j}\right)\right]+m_{2 i j} \log \left[\pi_{i j} \gamma_{j}\right]\right\}+\log C
, $$ where  $\boldsymbol{\pi_{i}} = \left(\pi_{i 1}, \ldots, \pi_{i J}\right)^T$, $\boldsymbol{\gamma} = \left(\gamma_{1}, \ldots, \gamma_{ J}\right)^T$, and $C=\prod_{j=1}^{J} \prod_{i=1}^{2} \frac{m_{+i j} !}{m_{0 i j} ! m_{1 i j} ! m_{2 i j} !}$ is a constant.

\subsection{Hypotheses for the proportion ratio across strata}

Assuming the ratio of proportions between two groups in the $j^{\text {th }}$ stratum is $\delta_{j}=\pi_{2 j} / \pi_{1 j}$ for $j=1, \ldots, J$, we construct three hypotheses as follows:

$$H_{}: \delta_{1}=\delta_{2}= ...=\delta_{J} \quad \text{(unconstrained hypothesis)}, $$
$$H_{}: \delta_{1}=\delta_{2}= ...=\delta_{J} = \delta_{0} \quad \text{(constrained hypothesis)}, $$
$$ H_{}: \delta_{1} \neq \delta_{2} \neq ... \neq \delta_{J} \quad \text{(global hypothesis)}.$$

Different hypotheses are based on different research questions: unconstrained hypothesis usually is used to test the homogeneity across strata; the constrained hypothesis indicates the common ratio of proportions by giving a constrained value of $\delta_{0}$; and the global hypothesis shows that none of $\delta$ are equal between strata.

\section{Maximum-likelihood estimation (MLE) under each hypothesis}

\subsection{The unconstrained MLEs}

Since we assume $\delta_{j}=\pi_{2 j} / \pi_{1 j}$ for $j=1, \ldots, J$, then $\pi_{2 j} = \delta_{j} \pi_{1 j}$, under the null hypothesis with common $\delta$, the log-likelihood ratio can be express as:

$$
l\left(\boldsymbol{\pi_1}, \delta, \boldsymbol{\gamma}\right)=\sum_{j=1}^J l_j\left(\pi_{1j}, \delta, \gamma_{j}\right),
$$
where 
$$
\begin{aligned}
\label{unconstrained_mle}
\quad l_{j}\left(\pi_{1 j}, \delta, \gamma_{j} \right)&= \{m_{01 j} \log \left[1-\left(2-\gamma_{j}\right) \pi_{1 j}\right] +m_{11 j} \log \left[2 \pi_{1 j}\left(1-\gamma_{j}\right)\right] +m_{21 j} \log \left[\pi_{1 j} \gamma_{j}\right]\\
&+m_{02 j} \log \left[1-\left(2-\gamma_{j}\right) \pi_{1 j} \delta\right]+m_{12 j} \log \left[2 \pi_{1 j} \delta\left(1-\gamma_{j}\right)\right] +m_{22 j} \log \left[\pi_{1 j} \delta \gamma_{j}\right] \}\\
&+\log C.
\end{aligned}
$$
Differentiating $ l\left(\boldsymbol{\pi_1}, \delta, \boldsymbol{\gamma}\right)$ with respect to $\boldsymbol{\pi_1}$ and $\boldsymbol{\gamma}$ , we have 

$$\frac{\partial^{} l_{}}{\partial \pi_{1 j}^{}} = \frac{ m_{1+j}}{ \pi_{1 j }} + \frac{ m_{2+j}}{ \pi_{1 j }} + \frac{ m_{01j}\,\left( \gamma_{ j}-2\right)}{ \pi_{1 j }\,\left( \gamma_{ j}-2\right)+1}+\frac{\delta \, m_{02j}\,\left( \gamma_{ j}-2\right)}{\delta \, \pi_{1 j }\,\left( \gamma_{ j}-2\right)+1},$$ 

$$\frac{\partial^{} l_{}}{\partial \gamma_{ j}^{}} = \frac{ m_{21j}}{ \gamma_{ j}}+\frac{ m_{22j}}{ \gamma_{ j}}+\frac{ m_{11j}}{ \gamma_{ j}-1}+\frac{ m_{12j}}{ \gamma_{ j}-1}+\frac{ m_{01j}\, \pi_{1 j }}{ \pi_{1 j }\,\left( \gamma_{ j}-2\right)+1}+\frac{\delta \, m_{02j}\, \pi_{1 j }}{\delta \, \pi_{1 j }\,\left( \gamma_{ j}-2\right)+1}.$$

We can set $\frac{\partial^{} l_{}}{\partial \pi_{1 j}^{}} =0 $ with $\frac{\partial^{} l_{}}{\partial \gamma_{ j}^{}} =0$ to get MLEs $\hat{\pi }_{1 j}$ and $\hat{\gamma}_j$. Indeed, MLE of $\gamma_{j}$ has closed-form solution. Meanwhile, MLE of $\pi_{1 j}$ is a function of $m$ and $\delta$.  
For the MLE of $\delta$, we will update by using Fisher scoring iterative algorithm, and 

$$\hat{\delta} = \delta^{(t+1)}=\delta^{(t)}+\left.I_{ }^{-1}\left(\delta^{(t)}\right)\left( \frac{\partial l_{ }}{\partial \delta}\right)\right|_{\delta=\delta(t), \hat{\pi }_{1 j}, \hat{\gamma_{j}} }, $$
where

$$\frac{\partial^{ } l_{}}{ \partial \delta } = \sum_{j=1}^{J} (\frac{ m_{12j} }{\delta }+\frac{ m_{22j} }{\delta }+\frac{ m_{02j} \, \pi_{1 j } \,{\left(\gamma_j -2\right)}}{\delta \, \pi_{1 j } \,{\left(\gamma_j -2\right)}+1}) , $$

$$
\frac{\partial^{2} l_{}}{ \partial \delta^{2} } =\sum_{j=1}^{J} (-\frac{ m_{12j} }{\delta^2 }-\frac{ m_{22j} }{\delta^2 }-\frac{ m_{02j} \,{ \pi_{1 j } }^2 \,{{\left(\gamma_j -2\right)}}^2 }{{{\left(\delta \, \pi_{1 j } \,{\left(\gamma_j -2\right)}+1\right)}}^2 }) ,
$$
and 
$$
\begin{aligned}
\label{I}
I = - E(\frac{\partial^{2} l_{}}{ \partial \delta^{2} }) &= -E (\sum_{j=1}^{J} (-\frac{ m_{12j} }{\delta^2 }-\frac{ m_{22j} }{\delta^2 }-\frac{ m_{02j} \,{ \pi_{1 j } }^2 \,{{\left(\gamma_j -2\right)}}^2 }{{{\left(\delta \, \pi_{1 j } \,{\left(\gamma_j -2\right)}+1\right)}}^2 })) \\
& =
\sum_{j=1}^{J} (\frac{ m_{+2j} p_{12j} }{\delta^2 }+\frac{ m_{+2j} p_{22j} }{\delta^2 }+\frac{ m_{+2j} p_{02j} \,{ \pi_{1 j } }^2 \,{{\left(\gamma_j -2\right)}}^2 }{{{\left(\delta \, \pi_{1 j } \,{\left(\gamma_j -2\right)}+1\right)}}^2 }) .
\end{aligned} 
$$

\subsection{The constrained MLEs}

The constrained hypothesis is nested within the unconstrained hypothesis, and the estimators of unconstrained MLEs are denoted as $\hat{\delta}$, $\hat{\pi }_{1 j}$, and $\hat{\gamma_{j}}$. Under the constrained hypothesis, we have $\pi_{2 j} = \delta_{0} \pi_{1 j}$ for all j. We can directly conclude that the constrained MLEs $\hat{\pi}_{1j_{H_{0}}}$, and $\hat{\gamma_{j}}_{H_{0}}$ have the same function as unconstrained MLEs by replacing $\delta$ with $\delta_{0}$.

\subsection{The Global MLEs}

Under the global hypothesis, $H_{a}: \delta_{1} \neq \delta_{2} \neq ... \neq \delta_{J}$ , the log-likelihood can be presented as:

$$
l\left(\boldsymbol{\pi_1}, \boldsymbol{\delta}, \boldsymbol{\gamma}\right)=\sum_{j=1}^J l_j\left(\pi_{1j}, \delta_{j}, \gamma_{j}\right),
$$
where
$$
\begin{aligned}
\label{globel_mle}
\quad l_{j}\left(\pi_{1 j}, \delta_{j}, \gamma_{j}\right) &=\sum_{j=1}^{J}\left\{m_{01 j} \log \left[1-\left(2-\gamma_{j}\right) \pi_{1 j}\right]\right. +m_{11 j} \log \left[2 \pi_{1 j}\left(1-\gamma_{j}\right)\right] +m_{21 j} \log \left[\pi_{1 j} \gamma_{j}\right] \\
&+m_{02 j} \log \left[1-\left(2-\gamma_{j}\right) \pi_{1 j} \delta_{j}\right]
+m_{12 j} \log \left[2 \pi_{1 j} \delta_{j}\left(1-\gamma_{j}\right)\right]
\left.+m_{22 j} \log \left[\pi_{1 j} \delta_{j} \gamma_{j}\right]\right\}+\log C ,
\end{aligned}
$$
and the MLEs of three parameters $\pi_{1 j}$, $\delta_{j}$, $\gamma_{j}$ can be derivied by setting the partial differentiation equal to zero, then the MLEs as following:

$$
\tilde {\pi}_{1 j} = \frac{{\left( m_{11j} + m_{21j} \right)}\,{\left( m_{11j} + m_{12j} +2\, m_{21j} +2\, m_{22j} \right)}}{2\,{\left( m_{01j} + m_{11j} + m_{21j} \right)}\,{\left( m_{11j} + m_{12j} + m_{21j} + m_{22j} \right)}} ,
$$

$$\tilde{\gamma}_{j} =\frac{2\, m_{2+j} }{ m_{1+j}+2\, m_{2+j} } , $$

$$\tilde{\delta_{j}} =  -\frac{{\left( m_{12j} + m_{22j} \right)}\,{\left( m_{01j} - m_{12j} + m_{1+j}- m_{22j} + m_{2+j} \right)}}{{\left( m_{02j} + m_{12j} + m_{22j} \right)}\,{\left( m_{12j} - m_{1+j}+ m_{22j} - m_{2+j} \right)}} . $$

\section{Confidence interval construction}

\subsection{Sample-size weighted global MLE-based Wald-type confidence interval (SGW)}

The sample-size weighted global MLE-based Wald-type confidence interval (SGW) is widely applied in stratified designs, and the idea of SGW places more weight on the stratum, resulting in more contribution to the overall sample \cite{Zhuang_2018CI, Xue_2019SMMR}. SGW is constructed by the global hypothesis: $ \delta_{1} \neq \delta_{2} \neq ... \neq \delta_{J}$, suppose that the MLEs of $\boldsymbol{\beta_j}=\left(\delta_j, \pi_{1 j}, \gamma_j\right)$ be $\boldsymbol{\tilde{\beta}_j}=\left(\tilde{\delta}_j, \tilde{\pi}_{1 j}, \tilde{\gamma}_j\right)$. Under regularity conditions, we have 

$$\sqrt{n}\left(\tilde{\boldsymbol{\beta}}_j^T-\boldsymbol{\beta}_j^T\right) \stackrel{d}{\rightarrow} N_3\left(\boldsymbol{0}, I_j^{-1}\right),$$ where $I_j$ is the Fisher information matrix of $\beta_j$. Let $c=(1,0,0)$, we can construct MLE of $\delta_{j}$ by $\tilde{\delta_{j}} = c\tilde{\boldsymbol{\beta}}_j^T$, and the asymptotic distribution of $\tilde{\delta_{j}}$ is:

$$
\sqrt{n}\left(\tilde{\delta}_j-\delta_j\right) \stackrel{d}{\rightarrow} N\left(0, c I_j^{-1} c^T\right) .
$$

Suppose the overall ratio of proportions is a linear combination of independent within stratum ratios of proportions, then the estimator $\tilde{\delta}=\sum_{j=1}^{\prime} w_j \tilde{\delta}_j$, where  $w_j$ 's are the weights for each stratum and restrict by $\sum_{j=1}^J w_j \delta_j=\delta, \quad \sum_{j=1}^J w_j=1.$ Therefore, the $100(1-\alpha) \%$ sample-size weighted global MLE-based Wald-type confidence limits of $\delta$ are defined as
$$
\sum_{j=1}^J w_j c \tilde{\boldsymbol{\beta}}_j^T \mp z_{1-\alpha / 2} \sqrt{\sum_{j=1}^J w_j^2 c I_j^{-1} c^T}
 , $$ where $z_{1-\alpha / 2}$ is the $100(1-\alpha / 2)$ percentile of the standard normal distribution.

By using the sample-size weight method, we assume $w_j=\frac{m_{++j}}{N}, \quad j=1, \ldots, J$, where $m_{++j}$ indicate the sample size of the $j^{\text {th }}$ stratum and $N$ present as the total sample size. Intuitively, the $100(1-\alpha) \%$ sample size weighted Wald-type confidence limits are 
$$
\begin{gathered}
\mathrm{LCL}_{SGW}=\max \left\{0, \sum_{j=1}^J \frac{m_{++j}}{N} c \tilde{\boldsymbol{\beta}}_j^T-z_{1-\alpha / 2} \sqrt{\sum_{j=1}^J\left(\frac{m_{++j}}{N}\right)^2 c I_j^{-1} c^T}\right\}, \\
\mathrm{UCL}_{SGW}=\sum_{j=1}^J \frac{m_{++j}}{N} c \tilde{\boldsymbol{\beta}}_j^T+z_{1-\alpha / 2} \sqrt{\sum_{j=1}^J\left(\frac{m_{++j}}{N}\right)^2 c I_j^{-1} c^T} . 
\end{gathered}
$$

\subsection{Complete MLE-based Wald-type confidence interval (CW)}

We consider another Wald-type confidence interval method, named complete MLE-based  Wald-type confidence interval (CW). This method is under unconstrained hypothesis: $ \delta_1=\delta_2=\cdots=\delta_J$, with considering homogeneity of proportion ratios. The MLEs of $\boldsymbol{\beta}=\left(\delta, \boldsymbol{\pi_{1}}, \boldsymbol{\gamma}\right)$ be $\hat{\boldsymbol{\beta}}=\left(\hat{\delta}_{}, \hat{\boldsymbol{\pi} }_{1 }, \hat{\boldsymbol{\gamma}}_{}\right)$. Under regularity conditions, we have
$$
\sqrt{n}\left(\hat{\boldsymbol{\beta}}^T- \boldsymbol{\beta}^T\right) \stackrel{d}{\rightarrow} N_{1+2 J}\left(0, I^{-1}\right) ,
$$
and $I$ corresponding to the fisher information matrix of $\beta$ (See Appendix for details). Assume $C=(1,0, \ldots, 0)_{1 \times(1+2 J)}$, the MLE of $\delta$ can be written as $\hat{\delta}=C \hat{\beta}^T$. By applying delta method, the asymptotic distribution of $\hat{\delta}$ is 
$$
\sqrt{n}(\hat{\delta}-\delta) \stackrel{d}{\rightarrow} N\left(0, C I^{-1} C^T\right).
$$

The $100(1-\alpha) \%$ of complete MLE-based Wald-type confidence limits are defined as
$$
\begin{gathered}
\mathrm{LCL}_{CW}=\max \left\{0, {C} \hat{\boldsymbol{\beta}}^T-z_{1-\alpha / 2} \sqrt{\left.{C} {I}^{-1} {C}^T\right\}},\right. \\
\mathrm{UCL}_{CW}=C \hat{\boldsymbol{\beta}}^T+z_{1-\alpha / 2} \sqrt{{C I ^ { - 1 }  { C } ^ { T }}} , 
\end{gathered} 
$$
where $z_{1-\alpha / 2}$ is the $100(1-\alpha / 2)$ percentile of the standard normal distribution.


\subsection{Profile likelihood confidence interval with the bisection algorithm (BI)}

The method of profile likelihood confidence interval with the bisection algorithm (BI) is constructed under hypothesis null hypothesis $\delta_1=\delta_2=\cdots=\delta_J=\delta_0$ (constrained hypothesis) vs. alternative hypothesis $\delta_1=\delta_2=\cdots=\delta_J$ (unconstrained hypothesis). Under regularity conditions, the likelihood ratio test statistic is given by

$$
\begin{aligned}
T_{L} &=2\left[l\left(\hat{\delta}, \hat{\boldsymbol{\pi} }_{1} , \hat{\boldsymbol{\gamma}}_{} \right)- l\left(\delta_0, \hat{\boldsymbol{\pi} }_{1{H_{0}}}, \hat{\boldsymbol{\gamma}}_{H_{0}}\right)\right].
\end{aligned}
$$

Under the null hypothesis, $T_L$ is asymptotically distributed as a Chi-square distribution with 1 degree of freedom. Therefore, the $100(1-\alpha) \%$ confidence interval for $\delta$ is written as

$$
C(\boldsymbol{m})=\left\{\delta: 2\left[l\left(\hat{\delta}_{}, \hat{\boldsymbol{\pi} }_{1} , \hat{\boldsymbol{\gamma}} \right)- l\left(\delta_0, \hat{\boldsymbol{\pi} }_{1{H_{0}}}, \hat{\boldsymbol{\gamma}}_{H_{0}}\right)\right]\leq \chi_{1,1-\alpha}^2\right\}.
$$

By bisection algorithm, we can get the estimates of the upper and lower limits of $\delta$. For the upper limit, the iteration follows: 

(1) Get the unconstrained MLEs $\hat{\delta}, \hat{\boldsymbol{\pi} }_{1}, \hat{\boldsymbol{\gamma}}$ first, then consider $\hat{\delta}$ as the initial value of $\delta$, set initial step size $s=0.1$ and direction indicator $d=1$.

(2) At the $t^{\text {th }}$ iteration, update $\hat{\delta}^{(t+1)}=\hat{\delta}^t+$  $ s\times d$.

(3) Given $\hat{\delta}^{(t+1)}$, construct the constrained MLEs $\left(\hat{{\boldsymbol{\pi}}}_{1 H_0}^{(t+1)}, \hat{\boldsymbol{\gamma}}_{H_0}^{(t+1)}\right)$.

(4) If $ d \times 2\left[l\left(\hat{\delta}, \hat{\boldsymbol{\pi} }_{1} , \hat{\boldsymbol{\gamma}} \right)-l\left(\hat{\delta}^{(t+1)}, \hat{{\boldsymbol{\pi}}}_{1 H_0}^{(t+1)}, \hat{\boldsymbol{\gamma}}_{H_0}^{(t+1)}\right)\right] \leq d \times \chi_{1,1-\alpha}^2$ return to step 2. Otherwise, $s=s\times 0.1$ and  $d =-d$, return to step 2.

(5) If the convergence is satisfactory and the length of step size is sufficiently small, then stop iterating and return $\hat{\delta}^{(t+1)}$ as the upper limit.

To obtain the lower limit, start with $d=-1$, and change evaluate the test statistic in step 4 to $ d \times 2\left[l\left(\hat{\delta}_{}, \hat{\boldsymbol{\pi} }_{1} , \hat{\boldsymbol{\gamma}} \right)-l\left(\hat{\delta}^{(t+1)}, \hat{{\boldsymbol{\pi}}}_{1 H_0}^{(t+1)}, \hat{\boldsymbol{\gamma}}_{H_0}^{(t+1)}\right)\right] \geq d \times \chi_{1,1-\alpha}^2 .$


\subsection{Complete MLE-based Score confidence interval (CS)}

The complete MLE-based score confidence interval (CS) is constructed by the constrained hypothesis $ \delta_1=\delta_2=\cdots=\delta_J=\delta_0$, the score test statistic is given by

$$
\begin{aligned}
T_{S C} &=\sum_{j=1}^{J} U_{j} I_{j}^{-1} U_{j}^{T} \mid \delta=\delta_{0}, \pi_{1 j}=\hat{\pi}_{1j{H_{0}}}, \gamma_{j}=\hat{\gamma}_{1j{H_{0}}} \\
&=\sum_{j=1}^{J}\left(\frac{\partial l_{j}}{\partial \delta}\right)^{2} I^{-1}_{(1,1) }\mid \delta_{ }=\delta_{0}, \pi_{1 j}=\hat{\pi}_{1j{H_{0}}}, \gamma_{j}=\hat{\gamma}_{1j{H_{0}}} ,
\end{aligned}
$$

where $U_{j}=\left(\frac{\partial l_{j}}{\partial \delta_{ }}, 0,0\right)$, and $I^{-1}_{(1,1) }$ is exactly the same as  complete MLE-based Wald-type confidence interval (more details are in the appendix).

Under the null hypothesis, $T_{S C}$ asymptotically follows a chi-square distribution with 1 degree of freedom, and
the $100(1-\alpha) \%$ Complete MLE-based Score confidence interval of $\delta$ that satisfies
$$
C(\boldsymbol{m})=\left\{\delta: T_{S C} \leq \chi_{1,1-\alpha}^2\right\},
$$
where $\chi_{1,1-\alpha}^2$ is the $100(1-\alpha) \%$ percentile of the $\chi^2$ distribution with 1 degree of freedom. By applying the bisection method with same procedure as section 4.3, the upper and lower limits can be directly constructed.

\section{Simulation study}

\subsection{Evaluation indices}

In this section, we evaluate the performance of the proposed four confidence interval estimators via Monte Carlo simulation studies in terms of three evaluation indices: the empirical mean coverage probabilities (MCPs), the empirical mean interval widths (MIWs), and the characterization index of an interval location (RMNCP). 

The MCP is constructed by the proportion of events that the true value of $\delta$ falls amid the simulated CIs, and the MCP is given by:

$$\mathrm{MCP}=\frac{1}{N} \sum_{n=1}^N I\left\{\delta \in\left[\delta_L\left(\boldsymbol{m}^{(n)}\right), \delta_U\left(\boldsymbol{m}^{(n)}\right)\right]\right\},$$
where $N$ denote the total number of replications, $\boldsymbol{m}^{(n)}$ represent the $n$th empirical samples, and the $\delta_L$ and $\delta_U$ indicate the estimates of the lower limit and upper limit of CIs, respectively. The remarkable thing is that MCP around $1- \alpha$ is preferred but considered conservative when MCP above $1- \alpha$. \cite{Zhuang_2018CI, newcombe1998improved, Xue_2019SMMR}. The MIW is defined as the sum of all constructed CIs widths divided by the total number of replications. By following the same notation as MCP, the MIW can be written as:

$$
\mathrm{MIW}=\frac{1}{N} \sum_{n=1}^N\left[\delta_U\left(\boldsymbol{m}^{(n)}\right)-\delta_L\left(\boldsymbol{m}^{(n)}\right)\right],
$$
and the smaller MIWs are recommend. RMNCP is described as mesial non-coverage  proportion of total non coverage, and we have:

$$ \text{RMNCP} =  \frac{\text{MNCP}}{\text{NCP}},$$
where NCP = MNCP + DNCP. The MNCP and DNCP are expressed as mesial and distal non coverage probability, respectively. Generally, MNCP and DNCP indicate the left-tail and right-tail error rates, where MNCP $=\sum_{n=1}^N I\left\{\delta<\delta_L\left(\boldsymbol{m}^{(n)}\right)\right\} / N$ and MNCP $=\sum_{n=1}^N I\left\{\delta<\delta_L\left(\boldsymbol{m}^{(n)}\right)\right\} / N$ \cite{newcombe1998improved, newcombe2011measures, newcombe1998two, newcombe2003confidence}. Therefore, RMNCP between 0.4 and 0.6 are recommended \cite{newcombe1998improved, newcombe2011measures}. 

\subsection{Simulation designs and results}

In the first Monte Carlo simulation study, we investigate the behavior of four proposed CIs by MCPs, MIWs, and  RMNCP under various procedures, where $m$ = 25, 50, or 100 in strata $J$ = 2, 4, 6, or 8. By considering that $\pi_{1 j}$ and $\gamma_{j}$ are common or different cross strata, we provide the parameter settings under different sample sizes and various sets of parameters in Table 1. 50,000 samples are randomly generated for each configuration under $\delta = 1, 1.2, 0.8$. As MCP showed in Table 2-17, we observe that the CIs of SGW are more conservative than the other three methods under small sample size ($m$ = 25) with multiple strata (J = 4, 6, 8) setting. However, as the sample size increases, SGW's CIs tend to be more satisfactory. Under all configurations, MIWs are shorter with sample size increase for four CI methods, and CIs of SGW provides wider MIWs than the other three methods. Meanwhile, BI and CS methods are more satisfactory in RMNCP under all configurations.        

\begin{singlespace}
\begin{longtable}[t]{llllll}
\caption{\label{tab:unnamed-chunk-6}Parameter setups}\\
\toprule
\multicolumn{1}{c}{} & \multicolumn{1}{c}{} & \multicolumn{4}{c}{Number of strata} \\
\cmidrule(l{3pt}r{3pt}){3-6}
Parameter & Cases & J=2 & J=4 & J=6 & J=8\\
\midrule
\endfirsthead
\caption[]{Parameter setups for computing empirical Type I error rates and powers. \textit{(continued)}}\\
\toprule
\multicolumn{1}{c}{} & \multicolumn{1}{c}{} & \multicolumn{4}{c}{Number of strata} \\
\cmidrule(l{3pt}r{3pt}){3-6}
Parameter & Cases & J=2 & J=4 & J=6 & J=8\\
\midrule
\endhead

\endfoot
\bottomrule
\endlastfoot
 & I & (0.2,0.4) & (0.2,0.4,0.2,0.4) & (0.2,0.4,0.2,0.4,0.2,0.4) & (0.2,0.4,0.2,0.4,0.2,0.4,0.2,0.4)\\
$\gamma$ & II & (0.3,0.3) & (0.3,0.3,0.3,0.3) & (0.3,0.3,0.3,0.3,0.3,0.3) & (0.3,0.3,0.3,0.3,0.3,0.3,0.3,0.3)\\
 & III & (0.3,0.5) & (0.3,0.5,0.3,0.5) & (0.3,0.5,0.3,0.5,0.3,0.5) & (0.3,0.5,0.3,0.5,0.3,0.5,0.3,0.5)\\
 & IV & (0.6,0.6) & (0.6,0.6,0.6,0.6) & (0.6,0.6,0.6,0.6,0.6,0.6) & (0.6,0.6,0.6,0.6,0.6,0.6,0.6,0.6)\\
 &  &  &  &  & \\

$\pi_{1}$ & a & (0.2,0.4) & (0.2,0.4,0.2,0.4) & (0.2,0.4,0.2,0.4,0.2,0.4) & (0.2,0.4,0.2,0.4,0.2,0.4,0.2,0.4)\\
 & b & (0.3,0.3) & (0.3,0.3,0.3,0.3) & (0.3,0.3,0.3,0.3,0.3,0.3) & (0.3,0.3,0.3,0.3,0.3,0.3,0.3,0.3)\\
 & c & (0.2,0.3) & (0.2,0.3,0.2,0.3) & (0.2,0.3,0.2,0.3,0.2,0.3) & (0.2,0.3,0.2,0.3,0.2,0.3,0.2,0.3)\\*
\end{longtable}
\end{singlespace}



\begin{singlespace}
\begin{longtable}[t]{rll>{}rrr>{}rrr>{}rrr}
\caption{\label{tab:unnamed-chunk-6}MCP, MIW and RMNCP of sample-size weighted Global MLE-based Wald-type confidence interval (SGW) for 2 strata.}\\
\toprule
\multicolumn{1}{c}{} & \multicolumn{1}{c}{} & \multicolumn{1}{c}{} & \multicolumn{3}{c}{m = 25} & \multicolumn{3}{c}{m = 50} & \multicolumn{3}{c}{m = 100} \\
\cmidrule(l{3pt}r{3pt}){4-6} \cmidrule(l{3pt}r{3pt}){7-9} \cmidrule(l{3pt}r{3pt}){10-12}
$\delta$ & $\gamma$ & $\pi_{1}$ &  MCP &  MIW &  RMNCP &  MCP &  MIW &  RMNCP &  MCP &  MIW &  RMNCP\\
\midrule
\endfirsthead
\caption[]{MCP, MIW and RMNCP of sample-size weighted Global MLE-based Wald-type confidence interval (SGW) for 2 strata. \textit{(continued)}}\\
\toprule
\multicolumn{1}{c}{} & \multicolumn{1}{c}{} & \multicolumn{1}{c}{} & \multicolumn{3}{c}{m = 25} & \multicolumn{3}{c}{m = 50} & \multicolumn{3}{c}{m = 100} \\
\cmidrule(l{3pt}r{3pt}){4-6} \cmidrule(l{3pt}r{3pt}){7-9} \cmidrule(l{3pt}r{3pt}){10-12}
$\delta$ & $\gamma$ & $\pi_{1}$ &  MCP &  MIW &  RMNCP &  MCP &  MIW &  RMNCP &  MCP &  MIW &  RMNCP\\
\midrule
\endhead

\endfoot
\bottomrule
\endlastfoot
\multicolumn{1}{l}{1.0} & \multicolumn{1}{l}{I} & \multicolumn{1}{l}{a} & \multicolumn{1}{l}{95.4} & \multicolumn{1}{l}{0.978} & \multicolumn{1}{l}{0.017} & \multicolumn{1}{l}{95.3} & \multicolumn{1}{l}{0.640} & \multicolumn{1}{l}{0.103} & \multicolumn{1}{l}{95.4} & \multicolumn{1}{l}{0.438} & \multicolumn{1}{l}{0.234}\\
\multicolumn{1}{l}{} & \multicolumn{1}{l}{} & \multicolumn{1}{l}{b} & \multicolumn{1}{l}{95.7} & \multicolumn{1}{l}{0.856} & \multicolumn{1}{l}{0.076} & \multicolumn{1}{l}{95.5} & \multicolumn{1}{l}{0.574} & \multicolumn{1}{l}{0.212} & \multicolumn{1}{l}{95.4} & \multicolumn{1}{l}{0.396} & \multicolumn{1}{l}{0.332}\\
\multicolumn{1}{l}{} & \multicolumn{1}{l}{} & \multicolumn{1}{l}{c} & \multicolumn{1}{l}{95.4} & \multicolumn{1}{l}{1.096} & \multicolumn{1}{l}{0.018} & \multicolumn{1}{l}{95.5} & \multicolumn{1}{l}{0.711} & \multicolumn{1}{l}{0.131} & \multicolumn{1}{l}{95.4} & \multicolumn{1}{l}{0.485} & \multicolumn{1}{l}{0.237}\\
\multicolumn{1}{l}{} & \multicolumn{1}{l}{II} & \multicolumn{1}{l}{a} & \multicolumn{1}{l}{94.9} & \multicolumn{1}{l}{1.007} & \multicolumn{1}{l}{0.005} & \multicolumn{1}{l}{95.6} & \multicolumn{1}{l}{0.655} & \multicolumn{1}{l}{0.067} & \multicolumn{1}{l}{95.3} & \multicolumn{1}{l}{0.445} & \multicolumn{1}{l}{0.195}\\
\multicolumn{1}{l}{} & \multicolumn{1}{l}{} & \multicolumn{1}{l}{b} & \multicolumn{1}{l}{95.8} & \multicolumn{1}{l}{0.852} & \multicolumn{1}{l}{0.077} & \multicolumn{1}{l}{95.4} & \multicolumn{1}{l}{0.571} & \multicolumn{1}{l}{0.228} & \multicolumn{1}{l}{95.2} & \multicolumn{1}{l}{0.393} & \multicolumn{1}{l}{0.330}\\

\multicolumn{1}{l}{} & \multicolumn{1}{l}{} & \multicolumn{1}{l}{c} & \multicolumn{1}{l}{95.3} & \multicolumn{1}{l}{1.115} & \multicolumn{1}{l}{0.014} & \multicolumn{1}{l}{95.4} & \multicolumn{1}{l}{0.719} & \multicolumn{1}{l}{0.100} & \multicolumn{1}{l}{95.4} & \multicolumn{1}{l}{0.489} & \multicolumn{1}{l}{0.223}\\
\end{longtable}
\end{singlespace}

\begin{singlespace}
\begin{longtable}[t]{rll>{}rrr>{}rrr>{}rrr}
\caption{\label{tab:unnamed-chunk-6}MCP, MIW and RMNCP of sample-size weighted Global MLE-based Wald-type confidence interval (SGW) for 4 strata.}\\
\toprule
\multicolumn{1}{c}{} & \multicolumn{1}{c}{} & \multicolumn{1}{c}{} & \multicolumn{3}{c}{m = 25} & \multicolumn{3}{c}{m = 50} & \multicolumn{3}{c}{m = 100} \\
\cmidrule(l{3pt}r{3pt}){4-6} \cmidrule(l{3pt}r{3pt}){7-9} \cmidrule(l{3pt}r{3pt}){10-12}
$\delta$ & $\gamma$ & $\pi_{1}$ &  MCP &  MIW &  RMNCP &  MCP &  MIW &  RMNCP &  MCP &  MIW &  RMNCP\\
\midrule
\endfirsthead
\caption[]{MCP, MIW and RMNCP of sample-size weighted Global MLE-based Wald-type confidence interval (SGW) for 4 strata. \textit{(continued)}}\\
\toprule
\multicolumn{1}{c}{} & \multicolumn{1}{c}{} & \multicolumn{1}{c}{} & \multicolumn{3}{c}{m = 25} & \multicolumn{3}{c}{m = 50} & \multicolumn{3}{c}{m = 100} \\
\cmidrule(l{3pt}r{3pt}){4-6} \cmidrule(l{3pt}r{3pt}){7-9} \cmidrule(l{3pt}r{3pt}){10-12}
$\delta$ & $\gamma$ & $\pi_{1}$ &  MCP &  MIW &  RMNCP &  MCP &  MIW &  RMNCP &  MCP &  MIW &  RMNCP\\
\midrule
\endhead

\endfoot
\bottomrule
\endlastfoot
\multicolumn{1}{l}{1.0} & \multicolumn{1}{l}{I} & \multicolumn{1}{l}{a} & \multicolumn{1}{l}{\textbf{96.6}} & \multicolumn{1}{l}{0.716} & \multicolumn{1}{l}{0.101} & \multicolumn{1}{l}{95.7} & \multicolumn{1}{l}{0.459} & \multicolumn{1}{l}{0.267} & \multicolumn{1}{l}{95.4} & \multicolumn{1}{l}{0.311} & \multicolumn{1}{l}{0.362}\\
\multicolumn{1}{l}{} & \multicolumn{1}{l}{} & \multicolumn{1}{l}{b} & \multicolumn{1}{l}{\textbf{96.3}} & \multicolumn{1}{l}{0.618} & \multicolumn{1}{l}{0.279} & \multicolumn{1}{l}{95.5} & \multicolumn{1}{l}{0.408} & \multicolumn{1}{l}{0.403} & \multicolumn{1}{l}{95.4} & \multicolumn{1}{l}{0.280} & \multicolumn{1}{l}{0.461}\\
\multicolumn{1}{l}{} & \multicolumn{1}{l}{} & \multicolumn{1}{l}{c} & \multicolumn{1}{l}{\textbf{96.7}} & \multicolumn{1}{l}{0.800} & \multicolumn{1}{l}{0.153} & \multicolumn{1}{l}{95.9} & \multicolumn{1}{l}{0.509} & \multicolumn{1}{l}{0.335} & \multicolumn{1}{l}{95.3} & \multicolumn{1}{l}{0.345} & \multicolumn{1}{l}{0.410}\\
\multicolumn{1}{l}{} & \multicolumn{1}{l}{II} & \multicolumn{1}{l}{a} & \multicolumn{1}{l}{\textbf{96.6}} & \multicolumn{1}{l}{0.742} & \multicolumn{1}{l}{0.070} & \multicolumn{1}{l}{\textbf{96.1}} & \multicolumn{1}{l}{0.469} & \multicolumn{1}{l}{0.210} & \multicolumn{1}{l}{95.6} & \multicolumn{1}{l}{0.317} & \multicolumn{1}{l}{0.359}\\
\multicolumn{1}{l}{} & \multicolumn{1}{l}{} & \multicolumn{1}{l}{b} & \multicolumn{1}{l}{\textbf{96.3}} & \multicolumn{1}{l}{0.614} & \multicolumn{1}{l}{0.282} & \multicolumn{1}{l}{95.4} & \multicolumn{1}{l}{0.406} & \multicolumn{1}{l}{0.395} & \multicolumn{1}{l}{95.3} & \multicolumn{1}{l}{0.279} & \multicolumn{1}{l}{0.446}\\

\multicolumn{1}{l}{} & \multicolumn{1}{l}{} & \multicolumn{1}{l}{c} & \multicolumn{1}{l}{\textbf{96.7}} & \multicolumn{1}{l}{0.816} & \multicolumn{1}{l}{0.129} & \multicolumn{1}{l}{\textbf{96.0}} & \multicolumn{1}{l}{0.516} & \multicolumn{1}{l}{0.286} & \multicolumn{1}{l}{95.5} & \multicolumn{1}{l}{0.348} & \multicolumn{1}{l}{0.384}\\
\end{longtable}
\end{singlespace}

\begin{singlespace}
\begin{longtable}[t]{rll>{}rrr>{}rrr>{}rrr}
\caption{\label{tab:unnamed-chunk-6}MCP, MIW and RMNCP of sample-size weighted Global MLE-based Wald-type confidence interval (SGW) for 6 strata.}\\
\toprule
\multicolumn{1}{c}{} & \multicolumn{1}{c}{} & \multicolumn{1}{c}{} & \multicolumn{3}{c}{m = 25} & \multicolumn{3}{c}{m = 50} & \multicolumn{3}{c}{m = 100} \\
\cmidrule(l{3pt}r{3pt}){4-6} \cmidrule(l{3pt}r{3pt}){7-9} \cmidrule(l{3pt}r{3pt}){10-12}
$\delta$ & $\gamma$ & $\pi_{1}$ &  MCP &  MIW &  RMNCP &  MCP &  MIW &  RMNCP &  MCP &  MIW &  RMNCP\\
\midrule
\endfirsthead
\caption[]{MCP, MIW and RMNCP of sample-size weighted Global MLE-based Wald-type confidence interval (SGW) for 6 strata. \textit{(continued)}}\\
\toprule
\multicolumn{1}{c}{} & \multicolumn{1}{c}{} & \multicolumn{1}{c}{} & \multicolumn{3}{c}{m = 25} & \multicolumn{3}{c}{m = 50} & \multicolumn{3}{c}{m = 100} \\
\cmidrule(l{3pt}r{3pt}){4-6} \cmidrule(l{3pt}r{3pt}){7-9} \cmidrule(l{3pt}r{3pt}){10-12}
$\delta$ & $\gamma$ & $\pi_{1}$ &  MCP &  MIW &  RMNCP &  MCP &  MIW &  RMNCP &  MCP &  MIW &  RMNCP\\
\midrule
\endhead

\endfoot
\bottomrule
\endlastfoot
\multicolumn{1}{l}{1.0} & \multicolumn{1}{l}{I} & \multicolumn{1}{l}{a} & \multicolumn{1}{l}{\textbf{97.0}} & \multicolumn{1}{l}{0.593} & \multicolumn{1}{l}{0.217} & \multicolumn{1}{l}{95.9} & \multicolumn{1}{l}{0.377} & \multicolumn{1}{l}{0.386} & \multicolumn{1}{l}{95.3} & \multicolumn{1}{l}{0.255} & \multicolumn{1}{l}{0.451}\\
\multicolumn{1}{l}{} & \multicolumn{1}{l}{} & \multicolumn{1}{l}{b} & \multicolumn{1}{l}{\textbf{96.1}} & \multicolumn{1}{l}{0.508} & \multicolumn{1}{l}{0.455} & \multicolumn{1}{l}{95.7} & \multicolumn{1}{l}{0.334} & \multicolumn{1}{l}{0.516} & \multicolumn{1}{l}{95.1} & \multicolumn{1}{l}{0.229} & \multicolumn{1}{l}{0.521}\\
\multicolumn{1}{l}{} & \multicolumn{1}{l}{} & \multicolumn{1}{l}{c} & \multicolumn{1}{l}{\textbf{97.0}} & \multicolumn{1}{l}{0.662} & \multicolumn{1}{l}{0.341} & \multicolumn{1}{l}{95.8} & \multicolumn{1}{l}{0.418} & \multicolumn{1}{l}{0.476} & \multicolumn{1}{l}{95.2} & \multicolumn{1}{l}{0.282} & \multicolumn{1}{l}{0.492}\\
\multicolumn{1}{l}{} & \multicolumn{1}{l}{II} & \multicolumn{1}{l}{a} & \multicolumn{1}{l}{\textbf{97.0}} & \multicolumn{1}{l}{0.618} & \multicolumn{1}{l}{0.156} & \multicolumn{1}{l}{\textbf{96.2}} & \multicolumn{1}{l}{0.386} & \multicolumn{1}{l}{0.358} & \multicolumn{1}{l}{95.5} & \multicolumn{1}{l}{0.260} & \multicolumn{1}{l}{0.429}\\
\multicolumn{1}{l}{} & \multicolumn{1}{l}{} & \multicolumn{1}{l}{b} & \multicolumn{1}{l}{\textbf{96.3}} & \multicolumn{1}{l}{0.505} & \multicolumn{1}{l}{0.461} & \multicolumn{1}{l}{95.6} & \multicolumn{1}{l}{0.333} & \multicolumn{1}{l}{0.520} & \multicolumn{1}{l}{95.2} & \multicolumn{1}{l}{0.228} & \multicolumn{1}{l}{0.537}\\

\multicolumn{1}{l}{} & \multicolumn{1}{l}{} & \multicolumn{1}{l}{c} & \multicolumn{1}{l}{\textbf{97.1}} & \multicolumn{1}{l}{0.678} & \multicolumn{1}{l}{0.277} & \multicolumn{1}{l}{\textbf{96.0}} & \multicolumn{1}{l}{0.424} & \multicolumn{1}{l}{0.451} & \multicolumn{1}{l}{95.4} & \multicolumn{1}{l}{0.285} & \multicolumn{1}{l}{0.481}\\
\end{longtable}
\end{singlespace}

\begin{singlespace}
\begin{longtable}[t]{rll>{}rrr>{}rrr>{}rrr}
\caption{\label{tab:unnamed-chunk-6}MCP, MIW and RMNCP of sample-size weighted Global MLE-based Wald-type confidence interval (SGW) for 8 strata.}\\
\toprule
\multicolumn{1}{c}{} & \multicolumn{1}{c}{} & \multicolumn{1}{c}{} & \multicolumn{3}{c}{m = 25} & \multicolumn{3}{c}{m = 50} & \multicolumn{3}{c}{m = 100} \\
\cmidrule(l{3pt}r{3pt}){4-6} \cmidrule(l{3pt}r{3pt}){7-9} \cmidrule(l{3pt}r{3pt}){10-12}
$\delta$ & $\gamma$ & $\pi_{1}$ &  MCP &  MIW &  RMNCP &  MCP &  MIW &  RMNCP &  MCP &  MIW &  RMNCP\\
\midrule
\endfirsthead
\caption[]{MCP, MIW and RMNCP of sample-size weighted Global MLE-based Wald-type confidence interval (SGW) for 8 strata. \textit{(continued)}}\\
\toprule
\multicolumn{1}{c}{} & \multicolumn{1}{c}{} & \multicolumn{1}{c}{} & \multicolumn{3}{c}{m = 25} & \multicolumn{3}{c}{m = 50} & \multicolumn{3}{c}{m = 100} \\
\cmidrule(l{3pt}r{3pt}){4-6} \cmidrule(l{3pt}r{3pt}){7-9} \cmidrule(l{3pt}r{3pt}){10-12}
$\delta$ & $\gamma$ & $\pi_{1}$ &  MCP &  MIW &  RMNCP &  MCP &  MIW &  RMNCP &  MCP &  MIW &  RMNCP\\
\midrule
\endhead

\endfoot
\bottomrule
\endlastfoot
\multicolumn{1}{l}{1.0} & \multicolumn{1}{l}{I} & \multicolumn{1}{l}{a} & \multicolumn{1}{l}{\textbf{97.1}} & \multicolumn{1}{l}{0.518} & \multicolumn{1}{l}{0.387} & \multicolumn{1}{l}{95.8} & \multicolumn{1}{l}{0.327} & \multicolumn{1}{l}{0.489} & \multicolumn{1}{l}{95.6} & \multicolumn{1}{l}{0.221} & \multicolumn{1}{l}{0.533}\\
\multicolumn{1}{l}{} & \multicolumn{1}{l}{} & \multicolumn{1}{l}{b} & \multicolumn{1}{l}{\textbf{96.1}} & \multicolumn{1}{l}{0.441} & \multicolumn{1}{l}{0.569} & \multicolumn{1}{l}{95.4} & \multicolumn{1}{l}{0.290} & \multicolumn{1}{l}{0.597} & \multicolumn{1}{l}{95.0} & \multicolumn{1}{l}{0.199} & \multicolumn{1}{l}{0.565}\\
\multicolumn{1}{l}{} & \multicolumn{1}{l}{} & \multicolumn{1}{l}{c} & \multicolumn{1}{l}{\textbf{96.8}} & \multicolumn{1}{l}{0.578} & \multicolumn{1}{l}{0.497} & \multicolumn{1}{l}{95.8} & \multicolumn{1}{l}{0.363} & \multicolumn{1}{l}{0.566} & \multicolumn{1}{l}{95.3} & \multicolumn{1}{l}{0.245} & \multicolumn{1}{l}{0.578}\\
\multicolumn{1}{l}{} & \multicolumn{1}{l}{II} & \multicolumn{1}{l}{a} & \multicolumn{1}{l}{\textbf{97.3}} & \multicolumn{1}{l}{0.542} & \multicolumn{1}{l}{0.302} & \multicolumn{1}{l}{\textbf{96.0}} & \multicolumn{1}{l}{0.335} & \multicolumn{1}{l}{0.456} & \multicolumn{1}{l}{95.3} & \multicolumn{1}{l}{0.225} & \multicolumn{1}{l}{0.521}\\
\multicolumn{1}{l}{} & \multicolumn{1}{l}{} & \multicolumn{1}{l}{b} & \multicolumn{1}{l}{\textbf{96.0}} & \multicolumn{1}{l}{0.439} & \multicolumn{1}{l}{0.573} & \multicolumn{1}{l}{95.3} & \multicolumn{1}{l}{0.289} & \multicolumn{1}{l}{0.606} & \multicolumn{1}{l}{95.1} & \multicolumn{1}{l}{0.198} & \multicolumn{1}{l}{0.564}\\

\multicolumn{1}{l}{} & \multicolumn{1}{l}{} & \multicolumn{1}{l}{c} & \multicolumn{1}{l}{\textbf{97.1}} & \multicolumn{1}{l}{0.593} & \multicolumn{1}{l}{0.448} & \multicolumn{1}{l}{95.8} & \multicolumn{1}{l}{0.368} & \multicolumn{1}{l}{0.559} & \multicolumn{1}{l}{95.3} & \multicolumn{1}{l}{0.247} & \multicolumn{1}{l}{0.574}\\
\end{longtable}
\end{singlespace}



\begin{singlespace}
\begin{longtable}[t]{rll>{}rrr>{}rrr>{}rrr}
\caption{\label{tab:unnamed-chunk-6}MCP, MIW and RMNCP of Complete MLE-based Wald-type confidence interval (CW) for 2 strata.}\\
\toprule
\multicolumn{1}{c}{} & \multicolumn{1}{c}{} & \multicolumn{1}{c}{} & \multicolumn{3}{c}{m = 25} & \multicolumn{3}{c}{m = 50} & \multicolumn{3}{c}{m = 100} \\
\cmidrule(l{3pt}r{3pt}){4-6} \cmidrule(l{3pt}r{3pt}){7-9} \cmidrule(l{3pt}r{3pt}){10-12}
$\delta$ & $\gamma$ & $\pi_{1}$ &  MCP &  MIW &  RMNCP &  MCP &  MIW &  RMNCP &  MCP &  MIW &  RMNCP\\
\midrule
\endfirsthead
\caption[]{MCP, MIW and RMNCP of Complete MLE-based Wald-type confidence interval (CW) for 2 strata. \textit{(continued)}}\\
\toprule
\multicolumn{1}{c}{} & \multicolumn{1}{c}{} & \multicolumn{1}{c}{} & \multicolumn{3}{c}{m = 25} & \multicolumn{3}{c}{m = 50} & \multicolumn{3}{c}{m = 100} \\
\cmidrule(l{3pt}r{3pt}){4-6} \cmidrule(l{3pt}r{3pt}){7-9} \cmidrule(l{3pt}r{3pt}){10-12}
$\delta$ & $\gamma$ & $\pi_{1}$ &  MCP &  MIW &  RMNCP &  MCP &  MIW &  RMNCP &  MCP &  MIW &  RMNCP\\
\midrule
\endhead

\endfoot
\bottomrule
\endlastfoot
\multicolumn{1}{l}{1.0} & \multicolumn{1}{l}{I} & \multicolumn{1}{l}{a} & \multicolumn{1}{l}{94.8} & \multicolumn{1}{l}{0.742} & \multicolumn{1}{l}{0.106} & \multicolumn{1}{l}{94.8} & \multicolumn{1}{l}{0.519} & \multicolumn{1}{l}{0.219} & \multicolumn{1}{l}{95.0} & \multicolumn{1}{l}{0.365} & \multicolumn{1}{l}{0.300}\\
\multicolumn{1}{l}{} & \multicolumn{1}{l}{} & \multicolumn{1}{l}{b} & \multicolumn{1}{l}{94.6} & \multicolumn{1}{l}{0.785} & \multicolumn{1}{l}{0.097} & \multicolumn{1}{l}{95.0} & \multicolumn{1}{l}{0.549} & \multicolumn{1}{l}{0.195} & \multicolumn{1}{l}{95.1} & \multicolumn{1}{l}{0.386} & \multicolumn{1}{l}{0.299}\\
\multicolumn{1}{l}{} & \multicolumn{1}{l}{} & \multicolumn{1}{l}{c} & \multicolumn{1}{l}{94.3} & \multicolumn{1}{l}{0.945} & \multicolumn{1}{l}{0.043} & \multicolumn{1}{l}{94.8} & \multicolumn{1}{l}{0.655} & \multicolumn{1}{l}{0.145} & \multicolumn{1}{l}{94.8} & \multicolumn{1}{l}{0.459} & \multicolumn{1}{l}{0.230}\\
\multicolumn{1}{l}{} & \multicolumn{1}{l}{II} & \multicolumn{1}{l}{a} & \multicolumn{1}{l}{94.9} & \multicolumn{1}{l}{0.698} & \multicolumn{1}{l}{0.101} & \multicolumn{1}{l}{95.1} & \multicolumn{1}{l}{0.489} & \multicolumn{1}{l}{0.238} & \multicolumn{1}{l}{94.9} & \multicolumn{1}{l}{0.343} & \multicolumn{1}{l}{0.309}\\
\multicolumn{1}{l}{} & \multicolumn{1}{l}{} & \multicolumn{1}{l}{b} & \multicolumn{1}{l}{94.7} & \multicolumn{1}{l}{0.787} & \multicolumn{1}{l}{0.092} & \multicolumn{1}{l}{94.8} & \multicolumn{1}{l}{0.550} & \multicolumn{1}{l}{0.204} & \multicolumn{1}{l}{94.9} & \multicolumn{1}{l}{0.387} & \multicolumn{1}{l}{0.291}\\

\multicolumn{1}{l}{} & \multicolumn{1}{l}{} & \multicolumn{1}{l}{c} & \multicolumn{1}{l}{94.3} & \multicolumn{1}{l}{0.924} & \multicolumn{1}{l}{0.045} & \multicolumn{1}{l}{94.8} & \multicolumn{1}{l}{0.640} & \multicolumn{1}{l}{0.146} & \multicolumn{1}{l}{95.0} & \multicolumn{1}{l}{0.448} & \multicolumn{1}{l}{0.242}\\
\end{longtable}
\end{singlespace}

\begin{singlespace}
\begin{longtable}[t]{rll>{}rrr>{}rrr>{}rrr}
\caption{\label{tab:unnamed-chunk-6}MCP, MIW and RMNCP of Complete MLE-based Wald-type confidence interval (CW) for 4 strata.}\\
\toprule
\multicolumn{1}{c}{} & \multicolumn{1}{c}{} & \multicolumn{1}{c}{} & \multicolumn{3}{c}{m = 25} & \multicolumn{3}{c}{m = 50} & \multicolumn{3}{c}{m = 100} \\
\cmidrule(l{3pt}r{3pt}){4-6} \cmidrule(l{3pt}r{3pt}){7-9} \cmidrule(l{3pt}r{3pt}){10-12}
$\delta$ & $\gamma$ & $\pi_{1}$ &  MCP &  MIW &  RMNCP &  MCP &  MIW &  RMNCP &  MCP &  MIW &  RMNCP\\
\midrule
\endfirsthead
\caption[]{MCP, MIW and RMNCP of Complete MLE-based Wald-type confidence interval (CW) for 4 strata. \textit{(continued)}}\\
\toprule
\multicolumn{1}{c}{} & \multicolumn{1}{c}{} & \multicolumn{1}{c}{} & \multicolumn{3}{c}{m = 25} & \multicolumn{3}{c}{m = 50} & \multicolumn{3}{c}{m = 100} \\
\cmidrule(l{3pt}r{3pt}){4-6} \cmidrule(l{3pt}r{3pt}){7-9} \cmidrule(l{3pt}r{3pt}){10-12}
$\delta$ & $\gamma$ & $\pi_{1}$ &  MCP &  MIW &  RMNCP &  MCP &  MIW &  RMNCP &  MCP &  MIW &  RMNCP\\
\midrule
\endhead

\endfoot
\bottomrule
\endlastfoot
\multicolumn{1}{l}{1.0} & \multicolumn{1}{l}{I} & \multicolumn{1}{l}{a} & \multicolumn{1}{l}{94.5} & \multicolumn{1}{l}{0.512} & \multicolumn{1}{l}{0.220} & \multicolumn{1}{l}{94.5} & \multicolumn{1}{l}{0.362} & \multicolumn{1}{l}{0.300} & \multicolumn{1}{l}{94.7} & \multicolumn{1}{l}{0.256} & \multicolumn{1}{l}{0.356}\\
\multicolumn{1}{l}{} & \multicolumn{1}{l}{} & \multicolumn{1}{l}{b} & \multicolumn{1}{l}{94.5} & \multicolumn{1}{l}{0.542} & \multicolumn{1}{l}{0.226} & \multicolumn{1}{l}{94.8} & \multicolumn{1}{l}{0.383} & \multicolumn{1}{l}{0.294} & \multicolumn{1}{l}{94.9} & \multicolumn{1}{l}{0.271} & \multicolumn{1}{l}{0.355}\\
\multicolumn{1}{l}{} & \multicolumn{1}{l}{} & \multicolumn{1}{l}{c} & \multicolumn{1}{l}{94.5} & \multicolumn{1}{l}{0.648} & \multicolumn{1}{l}{0.165} & \multicolumn{1}{l}{94.7} & \multicolumn{1}{l}{0.457} & \multicolumn{1}{l}{0.257} & \multicolumn{1}{l}{94.8} & \multicolumn{1}{l}{0.322} & \multicolumn{1}{l}{0.311}\\
\multicolumn{1}{l}{} & \multicolumn{1}{l}{II} & \multicolumn{1}{l}{a} & \multicolumn{1}{l}{94.5} & \multicolumn{1}{l}{0.481} & \multicolumn{1}{l}{0.249} & \multicolumn{1}{l}{94.7} & \multicolumn{1}{l}{0.340} & \multicolumn{1}{l}{0.293} & \multicolumn{1}{l}{95.1} & \multicolumn{1}{l}{0.241} & \multicolumn{1}{l}{0.366}\\
\multicolumn{1}{l}{} & \multicolumn{1}{l}{} & \multicolumn{1}{l}{b} & \multicolumn{1}{l}{94.5} & \multicolumn{1}{l}{0.544} & \multicolumn{1}{l}{0.209} & \multicolumn{1}{l}{94.5} & \multicolumn{1}{l}{0.384} & \multicolumn{1}{l}{0.280} & \multicolumn{1}{l}{94.9} & \multicolumn{1}{l}{0.272} & \multicolumn{1}{l}{0.339}\\

\multicolumn{1}{l}{} & \multicolumn{1}{l}{} & \multicolumn{1}{l}{c} & \multicolumn{1}{l}{94.5} & \multicolumn{1}{l}{0.634} & \multicolumn{1}{l}{0.168} & \multicolumn{1}{l}{94.8} & \multicolumn{1}{l}{0.446} & \multicolumn{1}{l}{0.230} & \multicolumn{1}{l}{94.9} & \multicolumn{1}{l}{0.315} & \multicolumn{1}{l}{0.315}\\
\end{longtable}
\end{singlespace}

\begin{singlespace}
\begin{longtable}[t]{rll>{}rrr>{}rrr>{}rrr}
\caption{\label{tab:unnamed-chunk-6}MCP, MIW and RMNCP of Complete MLE-based Wald-type confidence interval (CW) for 6 strata.}\\
\toprule
\multicolumn{1}{c}{} & \multicolumn{1}{c}{} & \multicolumn{1}{c}{} & \multicolumn{3}{c}{m = 25} & \multicolumn{3}{c}{m = 50} & \multicolumn{3}{c}{m = 100} \\
\cmidrule(l{3pt}r{3pt}){4-6} \cmidrule(l{3pt}r{3pt}){7-9} \cmidrule(l{3pt}r{3pt}){10-12}
$\delta$ & $\gamma$ & $\pi_{1}$ &  MCP &  MIW &  RMNCP &  MCP &  MIW &  RMNCP &  MCP &  MIW &  RMNCP\\
\midrule
\endfirsthead
\caption[]{MCP, MIW and RMNCP of Complete MLE-based Wald-type confidence interval (CW) for 6 strata. \textit{(continued)}}\\
\toprule
\multicolumn{1}{c}{} & \multicolumn{1}{c}{} & \multicolumn{1}{c}{} & \multicolumn{3}{c}{m = 25} & \multicolumn{3}{c}{m = 50} & \multicolumn{3}{c}{m = 100} \\
\cmidrule(l{3pt}r{3pt}){4-6} \cmidrule(l{3pt}r{3pt}){7-9} \cmidrule(l{3pt}r{3pt}){10-12}
$\delta$ & $\gamma$ & $\pi_{1}$ &  MCP &  MIW &  RMNCP &  MCP &  MIW &  RMNCP &  MCP &  MIW &  RMNCP\\
\midrule
\endhead

\endfoot
\bottomrule
\endlastfoot
\multicolumn{1}{l}{1.0} & \multicolumn{1}{l}{I} & \multicolumn{1}{l}{a} & \multicolumn{1}{l}{94.3} & \multicolumn{1}{l}{0.414} & \multicolumn{1}{l}{0.281} & \multicolumn{1}{l}{94.5} & \multicolumn{1}{l}{0.295} & \multicolumn{1}{l}{0.327} & \multicolumn{1}{l}{94.8} & \multicolumn{1}{l}{0.209} & \multicolumn{1}{l}{0.371}\\
\multicolumn{1}{l}{} & \multicolumn{1}{l}{} & \multicolumn{1}{l}{b} & \multicolumn{1}{l}{94.2} & \multicolumn{1}{l}{0.439} & \multicolumn{1}{l}{0.279} & \multicolumn{1}{l}{94.9} & \multicolumn{1}{l}{0.312} & \multicolumn{1}{l}{0.332} & \multicolumn{1}{l}{94.8} & \multicolumn{1}{l}{0.221} & \multicolumn{1}{l}{0.372}\\
\multicolumn{1}{l}{} & \multicolumn{1}{l}{} & \multicolumn{1}{l}{c} & \multicolumn{1}{l}{94.5} & \multicolumn{1}{l}{0.524} & \multicolumn{1}{l}{0.219} & \multicolumn{1}{l}{94.7} & \multicolumn{1}{l}{0.371} & \multicolumn{1}{l}{0.298} & \multicolumn{1}{l}{94.6} & \multicolumn{1}{l}{0.263} & \multicolumn{1}{l}{0.331}\\
\multicolumn{1}{l}{} & \multicolumn{1}{l}{II} & \multicolumn{1}{l}{a} & \multicolumn{1}{l}{94.3} & \multicolumn{1}{l}{0.389} & \multicolumn{1}{l}{0.284} & \multicolumn{1}{l}{94.7} & \multicolumn{1}{l}{0.277} & \multicolumn{1}{l}{0.339} & \multicolumn{1}{l}{94.8} & \multicolumn{1}{l}{0.196} & \multicolumn{1}{l}{0.390}\\
\multicolumn{1}{l}{} & \multicolumn{1}{l}{} & \multicolumn{1}{l}{b} & \multicolumn{1}{l}{94.4} & \multicolumn{1}{l}{0.441} & \multicolumn{1}{l}{0.263} & \multicolumn{1}{l}{94.8} & \multicolumn{1}{l}{0.313} & \multicolumn{1}{l}{0.326} & \multicolumn{1}{l}{94.9} & \multicolumn{1}{l}{0.222} & \multicolumn{1}{l}{0.377}\\

\multicolumn{1}{l}{} & \multicolumn{1}{l}{} & \multicolumn{1}{l}{c} & \multicolumn{1}{l}{94.5} & \multicolumn{1}{l}{0.512} & \multicolumn{1}{l}{0.223} & \multicolumn{1}{l}{94.7} & \multicolumn{1}{l}{0.363} & \multicolumn{1}{l}{0.298} & \multicolumn{1}{l}{94.7} & \multicolumn{1}{l}{0.256} & \multicolumn{1}{l}{0.345}\\
\end{longtable}
\end{singlespace}

\begin{singlespace}
\begin{longtable}[t]{rll>{}rrr>{}rrr>{}rrr}
\caption{\label{tab:unnamed-chunk-6}MCP, MIW and RMNCP of Complete MLE-based Wald-type confidence interval (CW) for 8 strata.}\\
\toprule
\multicolumn{1}{c}{} & \multicolumn{1}{c}{} & \multicolumn{1}{c}{} & \multicolumn{3}{c}{m = 25} & \multicolumn{3}{c}{m = 50} & \multicolumn{3}{c}{m = 100} \\
\cmidrule(l{3pt}r{3pt}){4-6} \cmidrule(l{3pt}r{3pt}){7-9} \cmidrule(l{3pt}r{3pt}){10-12}
$\delta$ & $\gamma$ & $\pi_{1}$ &  MCP &  MIW &  RMNCP &  MCP &  MIW &  RMNCP &  MCP &  MIW &  RMNCP\\
\midrule
\endfirsthead
\caption[]{MCP, MIW and RMNCP of Complete MLE-based Wald-type confidence interval (CW) for 8 strata. \textit{(continued)}}\\
\toprule
\multicolumn{1}{c}{} & \multicolumn{1}{c}{} & \multicolumn{1}{c}{} & \multicolumn{3}{c}{m = 25} & \multicolumn{3}{c}{m = 50} & \multicolumn{3}{c}{m = 100} \\
\cmidrule(l{3pt}r{3pt}){4-6} \cmidrule(l{3pt}r{3pt}){7-9} \cmidrule(l{3pt}r{3pt}){10-12}
$\delta$ & $\gamma$ & $\pi_{1}$ &  MCP &  MIW &  RMNCP &  MCP &  MIW &  RMNCP &  MCP &  MIW &  RMNCP\\
\midrule
\endhead

\endfoot
\bottomrule
\endlastfoot
\multicolumn{1}{l}{1.0} & \multicolumn{1}{l}{I} & \multicolumn{1}{l}{a} & \multicolumn{1}{l}{94.3} & \multicolumn{1}{l}{0.357} & \multicolumn{1}{l}{0.301} & \multicolumn{1}{l}{94.6} & \multicolumn{1}{l}{0.255} & \multicolumn{1}{l}{0.363} & \multicolumn{1}{l}{94.7} & \multicolumn{1}{l}{0.181} & \multicolumn{1}{l}{0.409}\\
\multicolumn{1}{l}{} & \multicolumn{1}{l}{} & \multicolumn{1}{l}{b} & \multicolumn{1}{l}{94.3} & \multicolumn{1}{l}{0.378} & \multicolumn{1}{l}{0.295} & \multicolumn{1}{l}{94.7} & \multicolumn{1}{l}{0.269} & \multicolumn{1}{l}{0.352} & \multicolumn{1}{l}{94.8} & \multicolumn{1}{l}{0.191} & \multicolumn{1}{l}{0.379}\\
\multicolumn{1}{l}{} & \multicolumn{1}{l}{} & \multicolumn{1}{l}{c} & \multicolumn{1}{l}{94.3} & \multicolumn{1}{l}{0.451} & \multicolumn{1}{l}{0.260} & \multicolumn{1}{l}{94.7} & \multicolumn{1}{l}{0.321} & \multicolumn{1}{l}{0.285} & \multicolumn{1}{l}{94.7} & \multicolumn{1}{l}{0.227} & \multicolumn{1}{l}{0.368}\\
\multicolumn{1}{l}{} & \multicolumn{1}{l}{II} & \multicolumn{1}{l}{a} & \multicolumn{1}{l}{94.2} & \multicolumn{1}{l}{0.335} & \multicolumn{1}{l}{0.323} & \multicolumn{1}{l}{94.6} & \multicolumn{1}{l}{0.239} & \multicolumn{1}{l}{0.351} & \multicolumn{1}{l}{94.8} & \multicolumn{1}{l}{0.170} & \multicolumn{1}{l}{0.400}\\
\multicolumn{1}{l}{} & \multicolumn{1}{l}{} & \multicolumn{1}{l}{b} & \multicolumn{1}{l}{94.4} & \multicolumn{1}{l}{0.380} & \multicolumn{1}{l}{0.294} & \multicolumn{1}{l}{94.6} & \multicolumn{1}{l}{0.270} & \multicolumn{1}{l}{0.354} & \multicolumn{1}{l}{94.8} & \multicolumn{1}{l}{0.192} & \multicolumn{1}{l}{0.387}\\

\multicolumn{1}{l}{} & \multicolumn{1}{l}{} & \multicolumn{1}{l}{c} & \multicolumn{1}{l}{94.3} & \multicolumn{1}{l}{0.441} & \multicolumn{1}{l}{0.261} & \multicolumn{1}{l}{94.7} & \multicolumn{1}{l}{0.314} & \multicolumn{1}{l}{0.320} & \multicolumn{1}{l}{94.9} & \multicolumn{1}{l}{0.222} & \multicolumn{1}{l}{0.390}\\
\end{longtable}
\end{singlespace}



\begin{singlespace}
\begin{longtable}[t]{rll>{}rrr>{}rrr>{}rrr}
\caption{\label{tab:unnamed-chunk-6}MCP, MIW and RMNCP of Profile likelihood confidence interval with the bisection algorithm (BI) for 2 strata.}\\
\toprule
\multicolumn{1}{c}{} & \multicolumn{1}{c}{} & \multicolumn{1}{c}{} & \multicolumn{3}{c}{m = 25} & \multicolumn{3}{c}{m = 50} & \multicolumn{3}{c}{m = 100} \\
\cmidrule(l{3pt}r{3pt}){4-6} \cmidrule(l{3pt}r{3pt}){7-9} \cmidrule(l{3pt}r{3pt}){10-12}
$\delta$ & $\gamma$ & $\pi_{1}$ &  MCP &  MIW &  RMNCP &  MCP &  MIW &  RMNCP &  MCP &  MIW &  RMNCP\\
\midrule
\endfirsthead
\caption[]{MCP, MIW and RMNCP of Profile likelihood confidence interval with the bisection algorithm (BI) for 2 strata. \textit{(continued)}}\\
\toprule
\multicolumn{1}{c}{} & \multicolumn{1}{c}{} & \multicolumn{1}{c}{} & \multicolumn{3}{c}{m = 25} & \multicolumn{3}{c}{m = 50} & \multicolumn{3}{c}{m = 100} \\
\cmidrule(l{3pt}r{3pt}){4-6} \cmidrule(l{3pt}r{3pt}){7-9} \cmidrule(l{3pt}r{3pt}){10-12}
$\delta$ & $\gamma$ & $\pi_{1}$ &  MCP &  MIW &  RMNCP &  MCP &  MIW &  RMNCP &  MCP &  MIW &  RMNCP\\
\midrule
\endhead

\endfoot
\bottomrule
\endlastfoot
\multicolumn{1}{l}{1.0} & \multicolumn{1}{l}{I} & \multicolumn{1}{l}{a} & \multicolumn{1}{l}{94.6} & \multicolumn{1}{l}{0.789} & \multicolumn{1}{l}{0.506} & \multicolumn{1}{l}{94.7} & \multicolumn{1}{l}{0.535} & \multicolumn{1}{l}{0.511} & \multicolumn{1}{l}{94.9} & \multicolumn{1}{l}{0.370} & \multicolumn{1}{l}{0.502}\\
\multicolumn{1}{l}{} & \multicolumn{1}{l}{} & \multicolumn{1}{l}{b} & \multicolumn{1}{l}{94.4} & \multicolumn{1}{l}{0.830} & \multicolumn{1}{l}{0.512} & \multicolumn{1}{l}{94.6} & \multicolumn{1}{l}{0.564} & \multicolumn{1}{l}{0.500} & \multicolumn{1}{l}{95.0} & \multicolumn{1}{l}{0.391} & \multicolumn{1}{l}{0.515}\\
\multicolumn{1}{l}{} & \multicolumn{1}{l}{} & \multicolumn{1}{l}{c} & \multicolumn{1}{l}{94.4} & \multicolumn{1}{l}{1.010} & \multicolumn{1}{l}{0.511} & \multicolumn{1}{l}{94.7} & \multicolumn{1}{l}{0.676} & \multicolumn{1}{l}{0.503} & \multicolumn{1}{l}{94.9} & \multicolumn{1}{l}{0.466} & \multicolumn{1}{l}{0.484}\\
\multicolumn{1}{l}{} & \multicolumn{1}{l}{II} & \multicolumn{1}{l}{a} & \multicolumn{1}{l}{94.5} & \multicolumn{1}{l}{0.744} & \multicolumn{1}{l}{0.499} & \multicolumn{1}{l}{94.9} & \multicolumn{1}{l}{0.505} & \multicolumn{1}{l}{0.511} & \multicolumn{1}{l}{94.8} & \multicolumn{1}{l}{0.349} & \multicolumn{1}{l}{0.508}\\
\multicolumn{1}{l}{} & \multicolumn{1}{l}{} & \multicolumn{1}{l}{b} & \multicolumn{1}{l}{94.6} & \multicolumn{1}{l}{0.832} & \multicolumn{1}{l}{0.501} & \multicolumn{1}{l}{94.5} & \multicolumn{1}{l}{0.565} & \multicolumn{1}{l}{0.508} & \multicolumn{1}{l}{94.9} & \multicolumn{1}{l}{0.392} & \multicolumn{1}{l}{0.510}\\

\multicolumn{1}{l}{} & \multicolumn{1}{l}{} & \multicolumn{1}{l}{c} & \multicolumn{1}{l}{94.4} & \multicolumn{1}{l}{0.988} & \multicolumn{1}{l}{0.502} & \multicolumn{1}{l}{94.9} & \multicolumn{1}{l}{0.661} & \multicolumn{1}{l}{0.490} & \multicolumn{1}{l}{95.0} & \multicolumn{1}{l}{0.455} & \multicolumn{1}{l}{0.501}\\
\end{longtable}
\end{singlespace}

\begin{singlespace}
\begin{longtable}[t]{rll>{}rrr>{}rrr>{}rrr}
\caption{\label{tab:unnamed-chunk-6}MCP, MIW and RMNCP of Profile likelihood confidence interval with the bisection algorithm (BI) for 4 strata.}\\
\toprule
\multicolumn{1}{c}{} & \multicolumn{1}{c}{} & \multicolumn{1}{c}{} & \multicolumn{3}{c}{m = 25} & \multicolumn{3}{c}{m = 50} & \multicolumn{3}{c}{m = 100} \\
\cmidrule(l{3pt}r{3pt}){4-6} \cmidrule(l{3pt}r{3pt}){7-9} \cmidrule(l{3pt}r{3pt}){10-12}
$\delta$ & $\gamma$ & $\pi_{1}$ &  MCP &  MIW &  RMNCP &  MCP &  MIW &  RMNCP &  MCP &  MIW &  RMNCP\\
\midrule
\endfirsthead
\caption[]{MCP, MIW and RMNCP of Profile likelihood confidence interval with the bisection algorithm (BI) for 4 strata. \textit{(continued)}}\\
\toprule
\multicolumn{1}{c}{} & \multicolumn{1}{c}{} & \multicolumn{1}{c}{} & \multicolumn{3}{c}{m = 25} & \multicolumn{3}{c}{m = 50} & \multicolumn{3}{c}{m = 100} \\
\cmidrule(l{3pt}r{3pt}){4-6} \cmidrule(l{3pt}r{3pt}){7-9} \cmidrule(l{3pt}r{3pt}){10-12}
$\delta$ & $\gamma$ & $\pi_{1}$ &  MCP &  MIW &  RMNCP &  MCP &  MIW &  RMNCP &  MCP &  MIW &  RMNCP\\
\midrule
\endhead

\endfoot
\bottomrule
\endlastfoot
\multicolumn{1}{l}{1.0} & \multicolumn{1}{l}{I} & \multicolumn{1}{l}{a} & \multicolumn{1}{l}{94.5} & \multicolumn{1}{l}{0.532} & \multicolumn{1}{l}{0.519} & \multicolumn{1}{l}{94.5} & \multicolumn{1}{l}{0.369} & \multicolumn{1}{l}{0.495} & \multicolumn{1}{l}{94.8} & \multicolumn{1}{l}{0.259} & \multicolumn{1}{l}{0.497}\\
\multicolumn{1}{l}{} & \multicolumn{1}{l}{} & \multicolumn{1}{l}{b} & \multicolumn{1}{l}{94.4} & \multicolumn{1}{l}{0.560} & \multicolumn{1}{l}{0.518} & \multicolumn{1}{l}{94.7} & \multicolumn{1}{l}{0.389} & \multicolumn{1}{l}{0.503} & \multicolumn{1}{l}{95.0} & \multicolumn{1}{l}{0.273} & \multicolumn{1}{l}{0.502}\\
\multicolumn{1}{l}{} & \multicolumn{1}{l}{} & \multicolumn{1}{l}{c} & \multicolumn{1}{l}{94.6} & \multicolumn{1}{l}{0.672} & \multicolumn{1}{l}{0.518} & \multicolumn{1}{l}{94.7} & \multicolumn{1}{l}{0.465} & \multicolumn{1}{l}{0.508} & \multicolumn{1}{l}{94.8} & \multicolumn{1}{l}{0.325} & \multicolumn{1}{l}{0.492}\\
\multicolumn{1}{l}{} & \multicolumn{1}{l}{II} & \multicolumn{1}{l}{a} & \multicolumn{1}{l}{94.4} & \multicolumn{1}{l}{0.501} & \multicolumn{1}{l}{0.509} & \multicolumn{1}{l}{94.7} & \multicolumn{1}{l}{0.347} & \multicolumn{1}{l}{0.483} & \multicolumn{1}{l}{95.1} & \multicolumn{1}{l}{0.244} & \multicolumn{1}{l}{0.522}\\
\multicolumn{1}{l}{} & \multicolumn{1}{l}{} & \multicolumn{1}{l}{b} & \multicolumn{1}{l}{94.6} & \multicolumn{1}{l}{0.562} & \multicolumn{1}{l}{0.504} & \multicolumn{1}{l}{94.6} & \multicolumn{1}{l}{0.391} & \multicolumn{1}{l}{0.488} & \multicolumn{1}{l}{94.9} & \multicolumn{1}{l}{0.274} & \multicolumn{1}{l}{0.499}\\

\multicolumn{1}{l}{} & \multicolumn{1}{l}{} & \multicolumn{1}{l}{c} & \multicolumn{1}{l}{94.5} & \multicolumn{1}{l}{0.657} & \multicolumn{1}{l}{0.503} & \multicolumn{1}{l}{94.8} & \multicolumn{1}{l}{0.454} & \multicolumn{1}{l}{0.485} & \multicolumn{1}{l}{94.9} & \multicolumn{1}{l}{0.318} & \multicolumn{1}{l}{0.495}\\
\end{longtable}
\end{singlespace}

\begin{singlespace}
\begin{longtable}[t]{rll>{}rrr>{}rrr>{}rrr}
\caption{\label{tab:unnamed-chunk-6}MCP, MIW and RMNCP of Profile likelihood confidence interval with the bisection algorithm (BI) for 6 strata.}\\
\toprule
\multicolumn{1}{c}{} & \multicolumn{1}{c}{} & \multicolumn{1}{c}{} & \multicolumn{3}{c}{m = 25} & \multicolumn{3}{c}{m = 50} & \multicolumn{3}{c}{m = 100} \\
\cmidrule(l{3pt}r{3pt}){4-6} \cmidrule(l{3pt}r{3pt}){7-9} \cmidrule(l{3pt}r{3pt}){10-12}
$\delta$ & $\gamma$ & $\pi_{1}$ &  MCP &  MIW &  RMNCP &  MCP &  MIW &  RMNCP &  MCP &  MIW &  RMNCP\\
\midrule
\endfirsthead
\caption[]{MCP, MIW and RMNCP of Profile likelihood confidence interval with the bisection algorithm (BI) for 6 strata. \textit{(continued)}}\\
\toprule
\multicolumn{1}{c}{} & \multicolumn{1}{c}{} & \multicolumn{1}{c}{} & \multicolumn{3}{c}{m = 25} & \multicolumn{3}{c}{m = 50} & \multicolumn{3}{c}{m = 100} \\
\cmidrule(l{3pt}r{3pt}){4-6} \cmidrule(l{3pt}r{3pt}){7-9} \cmidrule(l{3pt}r{3pt}){10-12}
$\delta$ & $\gamma$ & $\pi_{1}$ &  MCP &  MIW &  RMNCP &  MCP &  MIW &  RMNCP &  MCP &  MIW &  RMNCP\\
\midrule
\endhead

\endfoot
\bottomrule
\endlastfoot
\multicolumn{1}{l}{1.0} & \multicolumn{1}{l}{I} & \multicolumn{1}{l}{a} & \multicolumn{1}{l}{94.5} & \multicolumn{1}{l}{0.427} & \multicolumn{1}{l}{0.507} & \multicolumn{1}{l}{94.5} & \multicolumn{1}{l}{0.299} & \multicolumn{1}{l}{0.497} & \multicolumn{1}{l}{94.8} & \multicolumn{1}{l}{0.210} & \multicolumn{1}{l}{0.492}\\
\multicolumn{1}{l}{} & \multicolumn{1}{l}{} & \multicolumn{1}{l}{b} & \multicolumn{1}{l}{94.3} & \multicolumn{1}{l}{0.450} & \multicolumn{1}{l}{0.518} & \multicolumn{1}{l}{94.9} & \multicolumn{1}{l}{0.316} & \multicolumn{1}{l}{0.511} & \multicolumn{1}{l}{94.8} & \multicolumn{1}{l}{0.222} & \multicolumn{1}{l}{0.500}\\
\multicolumn{1}{l}{} & \multicolumn{1}{l}{} & \multicolumn{1}{l}{c} & \multicolumn{1}{l}{94.6} & \multicolumn{1}{l}{0.538} & \multicolumn{1}{l}{0.509} & \multicolumn{1}{l}{94.8} & \multicolumn{1}{l}{0.376} & \multicolumn{1}{l}{0.503} & \multicolumn{1}{l}{94.8} & \multicolumn{1}{l}{0.264} & \multicolumn{1}{l}{0.477}\\
\multicolumn{1}{l}{} & \multicolumn{1}{l}{II} & \multicolumn{1}{l}{a} & \multicolumn{1}{l}{94.4} & \multicolumn{1}{l}{0.403} & \multicolumn{1}{l}{0.503} & \multicolumn{1}{l}{94.8} & \multicolumn{1}{l}{0.282} & \multicolumn{1}{l}{0.483} & \multicolumn{1}{l}{94.9} & \multicolumn{1}{l}{0.198} & \multicolumn{1}{l}{0.503}\\
\multicolumn{1}{l}{} & \multicolumn{1}{l}{} & \multicolumn{1}{l}{b} & \multicolumn{1}{l}{94.7} & \multicolumn{1}{l}{0.452} & \multicolumn{1}{l}{0.502} & \multicolumn{1}{l}{94.9} & \multicolumn{1}{l}{0.317} & \multicolumn{1}{l}{0.505} & \multicolumn{1}{l}{94.9} & \multicolumn{1}{l}{0.223} & \multicolumn{1}{l}{0.504}\\

\multicolumn{1}{l}{} & \multicolumn{1}{l}{} & \multicolumn{1}{l}{c} & \multicolumn{1}{l}{94.4} & \multicolumn{1}{l}{0.526} & \multicolumn{1}{l}{0.497} & \multicolumn{1}{l}{94.8} & \multicolumn{1}{l}{0.368} & \multicolumn{1}{l}{0.509} & \multicolumn{1}{l}{94.8} & \multicolumn{1}{l}{0.258} & \multicolumn{1}{l}{0.484}\\
\end{longtable}

\end{singlespace}

\begin{singlespace}
\begin{longtable}[t]{rll>{}rrr>{}rrr>{}rrr}
\caption{\label{tab:unnamed-chunk-6}MCP, MIW and RMNCP of Profile likelihood confidence interval with the bisection algorithm (BI) for 8 strata.}\\
\toprule
\multicolumn{1}{c}{} & \multicolumn{1}{c}{} & \multicolumn{1}{c}{} & \multicolumn{3}{c}{m = 25} & \multicolumn{3}{c}{m = 50} & \multicolumn{3}{c}{m = 100} \\
\cmidrule(l{3pt}r{3pt}){4-6} \cmidrule(l{3pt}r{3pt}){7-9} \cmidrule(l{3pt}r{3pt}){10-12}
$\delta$ & $\gamma$ & $\pi_{1}$ &  MCP &  MIW &  RMNCP &  MCP &  MIW &  RMNCP &  MCP &  MIW &  RMNCP\\
\midrule
\endfirsthead
\caption[]{MCP, MIW and RMNCP of Profile likelihood confidence interval with the bisection algorithm (BI) for 8 strata. \textit{(continued)}}\\
\toprule
\multicolumn{1}{c}{} & \multicolumn{1}{c}{} & \multicolumn{1}{c}{} & \multicolumn{3}{c}{m = 25} & \multicolumn{3}{c}{m = 50} & \multicolumn{3}{c}{m = 100} \\
\cmidrule(l{3pt}r{3pt}){4-6} \cmidrule(l{3pt}r{3pt}){7-9} \cmidrule(l{3pt}r{3pt}){10-12}
$\delta$ & $\gamma$ & $\pi_{1}$ &  MCP &  MIW &  RMNCP &  MCP &  MIW &  RMNCP &  MCP &  MIW &  RMNCP\\
\midrule
\endhead

\endfoot
\bottomrule
\endlastfoot
\multicolumn{1}{l}{1.0} & \multicolumn{1}{l}{I} & \multicolumn{1}{l}{a} & \multicolumn{1}{l}{94.6} & \multicolumn{1}{l}{0.367} & \multicolumn{1}{l}{0.490} & \multicolumn{1}{l}{94.7} & \multicolumn{1}{l}{0.258} & \multicolumn{1}{l}{0.511} & \multicolumn{1}{l}{94.8} & \multicolumn{1}{l}{0.182} & \multicolumn{1}{l}{0.508}\\
\multicolumn{1}{l}{} & \multicolumn{1}{l}{} & \multicolumn{1}{l}{b} & \multicolumn{1}{l}{94.4} & \multicolumn{1}{l}{0.387} & \multicolumn{1}{l}{0.511} & \multicolumn{1}{l}{94.8} & \multicolumn{1}{l}{0.272} & \multicolumn{1}{l}{0.504} & \multicolumn{1}{l}{94.8} & \multicolumn{1}{l}{0.192} & \multicolumn{1}{l}{0.486}\\
\multicolumn{1}{l}{} & \multicolumn{1}{l}{} & \multicolumn{1}{l}{c} & \multicolumn{1}{l}{94.5} & \multicolumn{1}{l}{0.461} & \multicolumn{1}{l}{0.495} & \multicolumn{1}{l}{94.8} & \multicolumn{1}{l}{0.324} & \multicolumn{1}{l}{0.471} & \multicolumn{1}{l}{94.7} & \multicolumn{1}{l}{0.228} & \multicolumn{1}{l}{0.495}\\
\multicolumn{1}{l}{} & \multicolumn{1}{l}{II} & \multicolumn{1}{l}{a} & \multicolumn{1}{l}{94.3} & \multicolumn{1}{l}{0.346} & \multicolumn{1}{l}{0.504} & \multicolumn{1}{l}{94.7} & \multicolumn{1}{l}{0.243} & \multicolumn{1}{l}{0.488} & \multicolumn{1}{l}{94.8} & \multicolumn{1}{l}{0.171} & \multicolumn{1}{l}{0.500}\\
\multicolumn{1}{l}{} & \multicolumn{1}{l}{} & \multicolumn{1}{l}{b} & \multicolumn{1}{l}{94.6} & \multicolumn{1}{l}{0.389} & \multicolumn{1}{l}{0.499} & \multicolumn{1}{l}{94.8} & \multicolumn{1}{l}{0.273} & \multicolumn{1}{l}{0.509} & \multicolumn{1}{l}{94.9} & \multicolumn{1}{l}{0.193} & \multicolumn{1}{l}{0.488}\\

\multicolumn{1}{l}{} & \multicolumn{1}{l}{} & \multicolumn{1}{l}{c} & \multicolumn{1}{l}{94.4} & \multicolumn{1}{l}{0.451} & \multicolumn{1}{l}{0.489} & \multicolumn{1}{l}{94.8} & \multicolumn{1}{l}{0.317} & \multicolumn{1}{l}{0.494} & \multicolumn{1}{l}{94.9} & \multicolumn{1}{l}{0.223} & \multicolumn{1}{l}{0.515}\\
\end{longtable}

\end{singlespace}



\begin{singlespace}
\begin{longtable}[t]{rll>{}rrr>{}rrr>{}rrr}
\caption{\label{tab:unnamed-chunk-6}MCP, MIW and RMNCP of Complete MLE-based Score confidence interval (CS) for 2 strata.}\\
\toprule
\multicolumn{1}{c}{} & \multicolumn{1}{c}{} & \multicolumn{1}{c}{} & \multicolumn{3}{c}{m = 25} & \multicolumn{3}{c}{m = 50} & \multicolumn{3}{c}{m = 100} \\
\cmidrule(l{3pt}r{3pt}){4-6} \cmidrule(l{3pt}r{3pt}){7-9} \cmidrule(l{3pt}r{3pt}){10-12}
$\delta$ & $\gamma$ & $\pi_{1}$ &  MCP &  MIW &  RMNCP &  MCP &  MIW &  RMNCP &  MCP &  MIW &  RMNCP\\
\midrule
\endfirsthead
\caption[]{MCP, MIW and RMNCP of Complete MLE-based Score confidence interval (CS) for 2 strata. \textit{(continued)}}\\
\toprule
\multicolumn{1}{c}{} & \multicolumn{1}{c}{} & \multicolumn{1}{c}{} & \multicolumn{3}{c}{m = 25} & \multicolumn{3}{c}{m = 50} & \multicolumn{3}{c}{m = 100} \\
\cmidrule(l{3pt}r{3pt}){4-6} \cmidrule(l{3pt}r{3pt}){7-9} \cmidrule(l{3pt}r{3pt}){10-12}
$\delta$ & $\gamma$ & $\pi_{1}$ &  MCP &  MIW &  RMNCP &  MCP &  MIW &  RMNCP &  MCP &  MIW &  RMNCP\\
\midrule
\endhead

\endfoot
\bottomrule
\endlastfoot
\multicolumn{1}{l}{1.0} & \multicolumn{1}{l}{I} & \multicolumn{1}{l}{a} & \multicolumn{1}{l}{94.8} & \multicolumn{1}{l}{0.787} & \multicolumn{1}{l}{0.509} & \multicolumn{1}{l}{94.8} & \multicolumn{1}{l}{0.535} & \multicolumn{1}{l}{0.512} & \multicolumn{1}{l}{94.9} & \multicolumn{1}{l}{0.371} & \multicolumn{1}{l}{0.503}\\
\multicolumn{1}{l}{} & \multicolumn{1}{l}{} & \multicolumn{1}{l}{b} & \multicolumn{1}{l}{94.5} & \multicolumn{1}{l}{0.823} & \multicolumn{1}{l}{0.511} & \multicolumn{1}{l}{94.8} & \multicolumn{1}{l}{0.562} & \multicolumn{1}{l}{0.502} & \multicolumn{1}{l}{95.1} & \multicolumn{1}{l}{0.390} & \multicolumn{1}{l}{0.514}\\
\multicolumn{1}{l}{} & \multicolumn{1}{l}{} & \multicolumn{1}{l}{c} & \multicolumn{1}{l}{94.6} & \multicolumn{1}{l}{0.993} & \multicolumn{1}{l}{0.509} & \multicolumn{1}{l}{94.8} & \multicolumn{1}{l}{0.671} & \multicolumn{1}{l}{0.505} & \multicolumn{1}{l}{94.9} & \multicolumn{1}{l}{0.464} & \multicolumn{1}{l}{0.484}\\
\multicolumn{1}{l}{} & \multicolumn{1}{l}{II} & \multicolumn{1}{l}{a} & \multicolumn{1}{l}{94.8} & \multicolumn{1}{l}{0.746} & \multicolumn{1}{l}{0.497} & \multicolumn{1}{l}{95.0} & \multicolumn{1}{l}{0.506} & \multicolumn{1}{l}{0.512} & \multicolumn{1}{l}{94.8} & \multicolumn{1}{l}{0.349} & \multicolumn{1}{l}{0.508}\\
\multicolumn{1}{l}{} & \multicolumn{1}{l}{} & \multicolumn{1}{l}{b} & \multicolumn{1}{l}{94.8} & \multicolumn{1}{l}{0.824} & \multicolumn{1}{l}{0.500} & \multicolumn{1}{l}{94.6} & \multicolumn{1}{l}{0.564} & \multicolumn{1}{l}{0.506} & \multicolumn{1}{l}{95.0} & \multicolumn{1}{l}{0.391} & \multicolumn{1}{l}{0.509}\\

\multicolumn{1}{l}{} & \multicolumn{1}{l}{} & \multicolumn{1}{l}{c} & \multicolumn{1}{l}{94.6} & \multicolumn{1}{l}{0.972} & \multicolumn{1}{l}{0.504} & \multicolumn{1}{l}{94.9} & \multicolumn{1}{l}{0.656} & \multicolumn{1}{l}{0.489} & \multicolumn{1}{l}{95.0} & \multicolumn{1}{l}{0.454} & \multicolumn{1}{l}{0.502}\\
\end{longtable}
\end{singlespace}


\begin{singlespace}
\begin{longtable}[t]{rll>{}rrr>{}rrr>{}rrr}
\caption{\label{tab:unnamed-chunk-6}MCP, MIW and RMNCP of Complete MLE-based Score confidence interval (CS) for 4 strata.}\\
\toprule
\multicolumn{1}{c}{} & \multicolumn{1}{c}{} & \multicolumn{1}{c}{} & \multicolumn{3}{c}{m = 25} & \multicolumn{3}{c}{m = 50} & \multicolumn{3}{c}{m = 100} \\
\cmidrule(l{3pt}r{3pt}){4-6} \cmidrule(l{3pt}r{3pt}){7-9} \cmidrule(l{3pt}r{3pt}){10-12}
$\delta$ & $\gamma$ & $\pi_{1}$ &  MCP &  MIW &  RMNCP &  MCP &  MIW &  RMNCP &  MCP &  MIW &  RMNCP\\
\midrule
\endfirsthead
\caption[]{MCP, MIW and RMNCP of Complete MLE-based Score confidence interval (CS) for 4 strata. \textit{(continued)}}\\
\toprule
\multicolumn{1}{c}{} & \multicolumn{1}{c}{} & \multicolumn{1}{c}{} & \multicolumn{3}{c}{m = 25} & \multicolumn{3}{c}{m = 50} & \multicolumn{3}{c}{m = 100} \\
\cmidrule(l{3pt}r{3pt}){4-6} \cmidrule(l{3pt}r{3pt}){7-9} \cmidrule(l{3pt}r{3pt}){10-12}
$\delta$ & $\gamma$ & $\pi_{1}$ &  MCP &  MIW &  RMNCP &  MCP &  MIW &  RMNCP &  MCP &  MIW &  RMNCP\\
\midrule
\endhead

\endfoot
\bottomrule
\endlastfoot
\multicolumn{1}{l}{1.0} & \multicolumn{1}{l}{I} & \multicolumn{1}{l}{a} & \multicolumn{1}{l}{94.8} & \multicolumn{1}{l}{0.535} & \multicolumn{1}{l}{0.523} & \multicolumn{1}{l}{94.7} & \multicolumn{1}{l}{0.371} & \multicolumn{1}{l}{0.497} & \multicolumn{1}{l}{94.9} & \multicolumn{1}{l}{0.259} & \multicolumn{1}{l}{0.498}\\
\multicolumn{1}{l}{} & \multicolumn{1}{l}{} & \multicolumn{1}{l}{b} & \multicolumn{1}{l}{94.7} & \multicolumn{1}{l}{0.562} & \multicolumn{1}{l}{0.517} & \multicolumn{1}{l}{94.9} & \multicolumn{1}{l}{0.390} & \multicolumn{1}{l}{0.504} & \multicolumn{1}{l}{95.1} & \multicolumn{1}{l}{0.273} & \multicolumn{1}{l}{0.502}\\
\multicolumn{1}{l}{} & \multicolumn{1}{l}{} & \multicolumn{1}{l}{c} & \multicolumn{1}{l}{94.8} & \multicolumn{1}{l}{0.670} & \multicolumn{1}{l}{0.515} & \multicolumn{1}{l}{94.8} & \multicolumn{1}{l}{0.464} & \multicolumn{1}{l}{0.508} & \multicolumn{1}{l}{94.9} & \multicolumn{1}{l}{0.325} & \multicolumn{1}{l}{0.493}\\
\multicolumn{1}{l}{} & \multicolumn{1}{l}{II} & \multicolumn{1}{l}{a} & \multicolumn{1}{l}{94.9} & \multicolumn{1}{l}{0.507} & \multicolumn{1}{l}{0.510} & \multicolumn{1}{l}{94.9} & \multicolumn{1}{l}{0.350} & \multicolumn{1}{l}{0.483} & \multicolumn{1}{l}{95.1} & \multicolumn{1}{l}{0.244} & \multicolumn{1}{l}{0.521}\\
\multicolumn{1}{l}{} & \multicolumn{1}{l}{} & \multicolumn{1}{l}{b} & \multicolumn{1}{l}{94.9} & \multicolumn{1}{l}{0.563} & \multicolumn{1}{l}{0.504} & \multicolumn{1}{l}{94.7} & \multicolumn{1}{l}{0.391} & \multicolumn{1}{l}{0.489} & \multicolumn{1}{l}{95.0} & \multicolumn{1}{l}{0.274} & \multicolumn{1}{l}{0.501}\\

\multicolumn{1}{l}{} & \multicolumn{1}{l}{} & \multicolumn{1}{l}{c} & \multicolumn{1}{l}{94.7} & \multicolumn{1}{l}{0.656} & \multicolumn{1}{l}{0.506} & \multicolumn{1}{l}{94.9} & \multicolumn{1}{l}{0.454} & \multicolumn{1}{l}{0.483} & \multicolumn{1}{l}{94.9} & \multicolumn{1}{l}{0.318} & \multicolumn{1}{l}{0.495}\\
\end{longtable}
\end{singlespace}

\begin{singlespace}
\begin{longtable}[t]{rll>{}rrr>{}rrr>{}rrr}
\caption{\label{tab:unnamed-chunk-6}MCP, MIW and RMNCP of Complete MLE-based Score confidence interval (CS) for 6 strata.}\\
\toprule
\multicolumn{1}{c}{} & \multicolumn{1}{c}{} & \multicolumn{1}{c}{} & \multicolumn{3}{c}{m = 25} & \multicolumn{3}{c}{m = 50} & \multicolumn{3}{c}{m = 100} \\
\cmidrule(l{3pt}r{3pt}){4-6} \cmidrule(l{3pt}r{3pt}){7-9} \cmidrule(l{3pt}r{3pt}){10-12}
$\delta$ & $\gamma$ & $\pi_{1}$ &  MCP &  MIW &  RMNCP &  MCP &  MIW &  RMNCP &  MCP &  MIW &  RMNCP\\
\midrule
\endfirsthead
\caption[]{MCP, MIW and RMNCP of Complete MLE-based Score confidence interval (CS) for 6 strata. \textit{(continued)}}\\
\toprule
\multicolumn{1}{c}{} & \multicolumn{1}{c}{} & \multicolumn{1}{c}{} & \multicolumn{3}{c}{m = 25} & \multicolumn{3}{c}{m = 50} & \multicolumn{3}{c}{m = 100} \\
\cmidrule(l{3pt}r{3pt}){4-6} \cmidrule(l{3pt}r{3pt}){7-9} \cmidrule(l{3pt}r{3pt}){10-12}
$\delta$ & $\gamma$ & $\pi_{1}$ &  MCP &  MIW &  RMNCP &  MCP &  MIW &  RMNCP &  MCP &  MIW &  RMNCP\\
\midrule
\endhead

\endfoot
\bottomrule
\endlastfoot
\multicolumn{1}{l}{1.0} & \multicolumn{1}{l}{I} & \multicolumn{1}{l}{a} & \multicolumn{1}{l}{94.9} & \multicolumn{1}{l}{0.431} & \multicolumn{1}{l}{0.507} & \multicolumn{1}{l}{94.7} & \multicolumn{1}{l}{0.301} & \multicolumn{1}{l}{0.496} & \multicolumn{1}{l}{94.9} & \multicolumn{1}{l}{0.211} & \multicolumn{1}{l}{0.493}\\
\multicolumn{1}{l}{} & \multicolumn{1}{l}{} & \multicolumn{1}{l}{b} & \multicolumn{1}{l}{94.6} & \multicolumn{1}{l}{0.453} & \multicolumn{1}{l}{0.520} & \multicolumn{1}{l}{95.1} & \multicolumn{1}{l}{0.316} & \multicolumn{1}{l}{0.514} & \multicolumn{1}{l}{94.9} & \multicolumn{1}{l}{0.222} & \multicolumn{1}{l}{0.500}\\
\multicolumn{1}{l}{} & \multicolumn{1}{l}{} & \multicolumn{1}{l}{c} & \multicolumn{1}{l}{94.8} & \multicolumn{1}{l}{0.538} & \multicolumn{1}{l}{0.508} & \multicolumn{1}{l}{94.9} & \multicolumn{1}{l}{0.376} & \multicolumn{1}{l}{0.505} & \multicolumn{1}{l}{94.9} & \multicolumn{1}{l}{0.264} & \multicolumn{1}{l}{0.476}\\
\multicolumn{1}{l}{} & \multicolumn{1}{l}{II} & \multicolumn{1}{l}{a} & \multicolumn{1}{l}{94.9} & \multicolumn{1}{l}{0.408} & \multicolumn{1}{l}{0.502} & \multicolumn{1}{l}{95.0} & \multicolumn{1}{l}{0.284} & \multicolumn{1}{l}{0.484} & \multicolumn{1}{l}{95.0} & \multicolumn{1}{l}{0.199} & \multicolumn{1}{l}{0.504}\\
\multicolumn{1}{l}{} & \multicolumn{1}{l}{} & \multicolumn{1}{l}{b} & \multicolumn{1}{l}{94.9} & \multicolumn{1}{l}{0.454} & \multicolumn{1}{l}{0.504} & \multicolumn{1}{l}{95.0} & \multicolumn{1}{l}{0.318} & \multicolumn{1}{l}{0.508} & \multicolumn{1}{l}{95.0} & \multicolumn{1}{l}{0.223} & \multicolumn{1}{l}{0.504}\\

\multicolumn{1}{l}{} & \multicolumn{1}{l}{} & \multicolumn{1}{l}{c} & \multicolumn{1}{l}{94.6} & \multicolumn{1}{l}{0.527} & \multicolumn{1}{l}{0.496} & \multicolumn{1}{l}{94.9} & \multicolumn{1}{l}{0.368} & \multicolumn{1}{l}{0.508} & \multicolumn{1}{l}{94.8} & \multicolumn{1}{l}{0.258} & \multicolumn{1}{l}{0.483}\\
\end{longtable}%
\end{singlespace}

To obtain completed and robust empirical performances of four proposed CIs, we propose the second simulation by randomly choosing parameter configurations. We randomly generate 1,000 parameter configurations within parameter space for different sample sizes $m$ = 25, 50, 100, and variety of strata $J$ = 2, 4, 6, and 8 for 50,000 replications. Furthermore, the corresponding boxplots and violin plots are shown in Figures 1 to 9. Same as the first simulation, the MCPs of CIs in the SGW method are more conservative than the other three methods under the small sample size ($m$ = 25) with multiple strata (J = 4, 6, 8) setting. Meanwhile, the CS method always leads to the best performance of MCPs under all configurations. MIWs of the GSW method has wider CIs under most configurations. Under all configurations, BI and CS methods produce relatively satisfactory RMNCP.

\begin{figure}[htp]
    \centering
    \includegraphics[width=16cm]{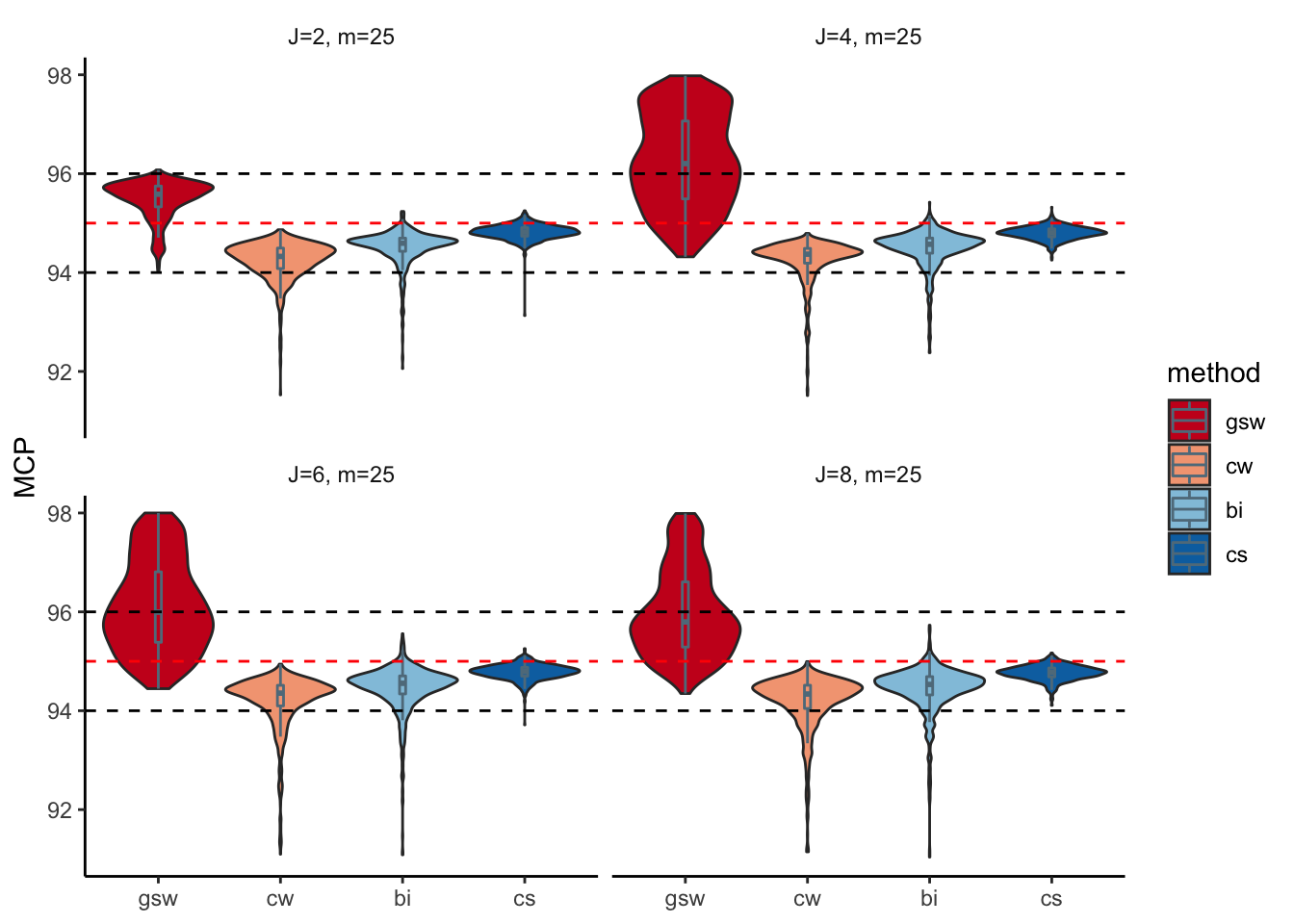}
    \caption{Violin-plots and Boxplots of MCP with m=25}
    \label{fig:galaxy}
\end{figure}

\begin{figure}[htp]
    \centering
    \includegraphics[width=16cm]{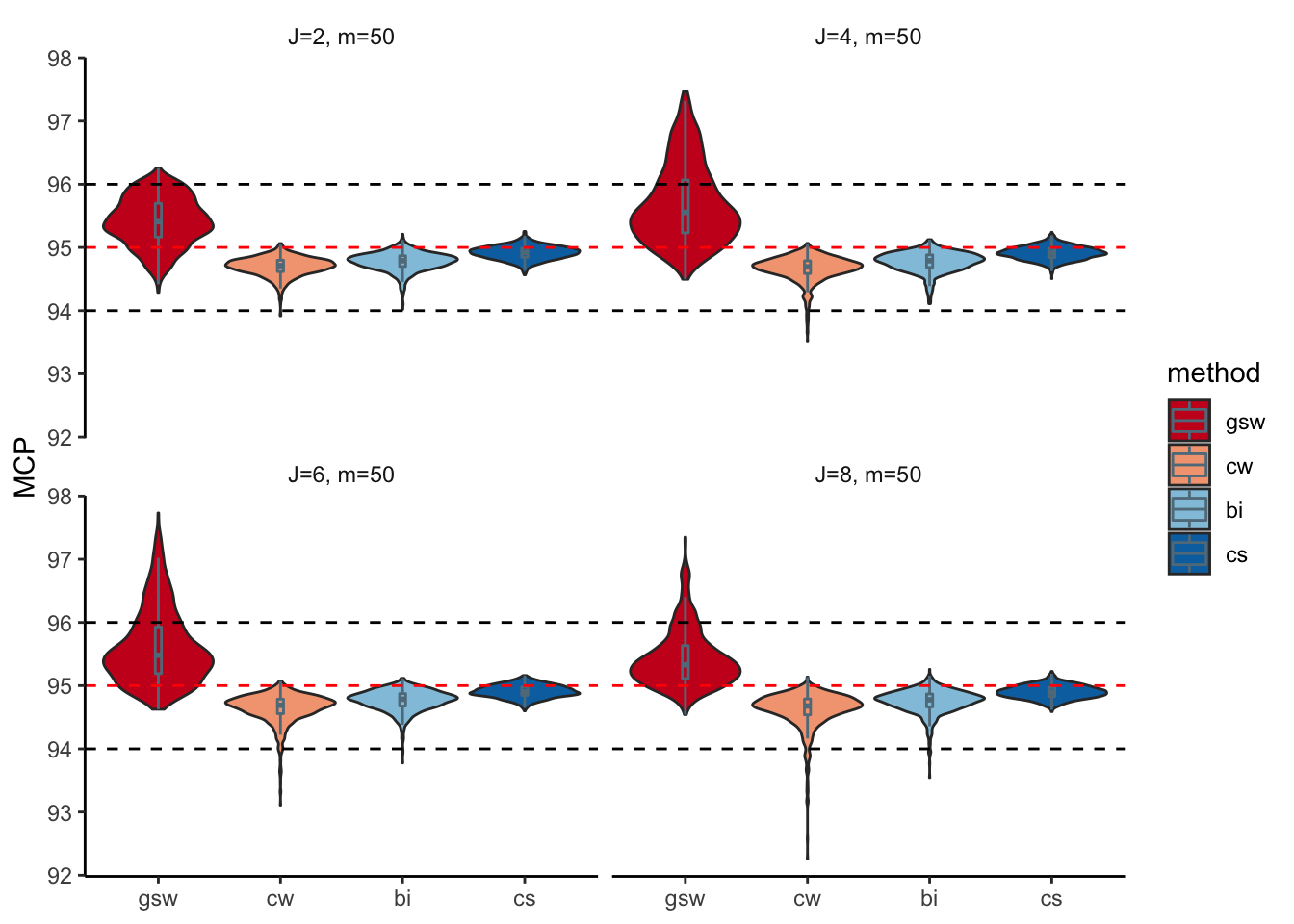}
    \caption{Violin-plots and Boxplots of MCP with m=50}
    \label{fig:galaxy}
\end{figure}

\begin{figure}[htp]
    \centering
    \includegraphics[width=16cm]{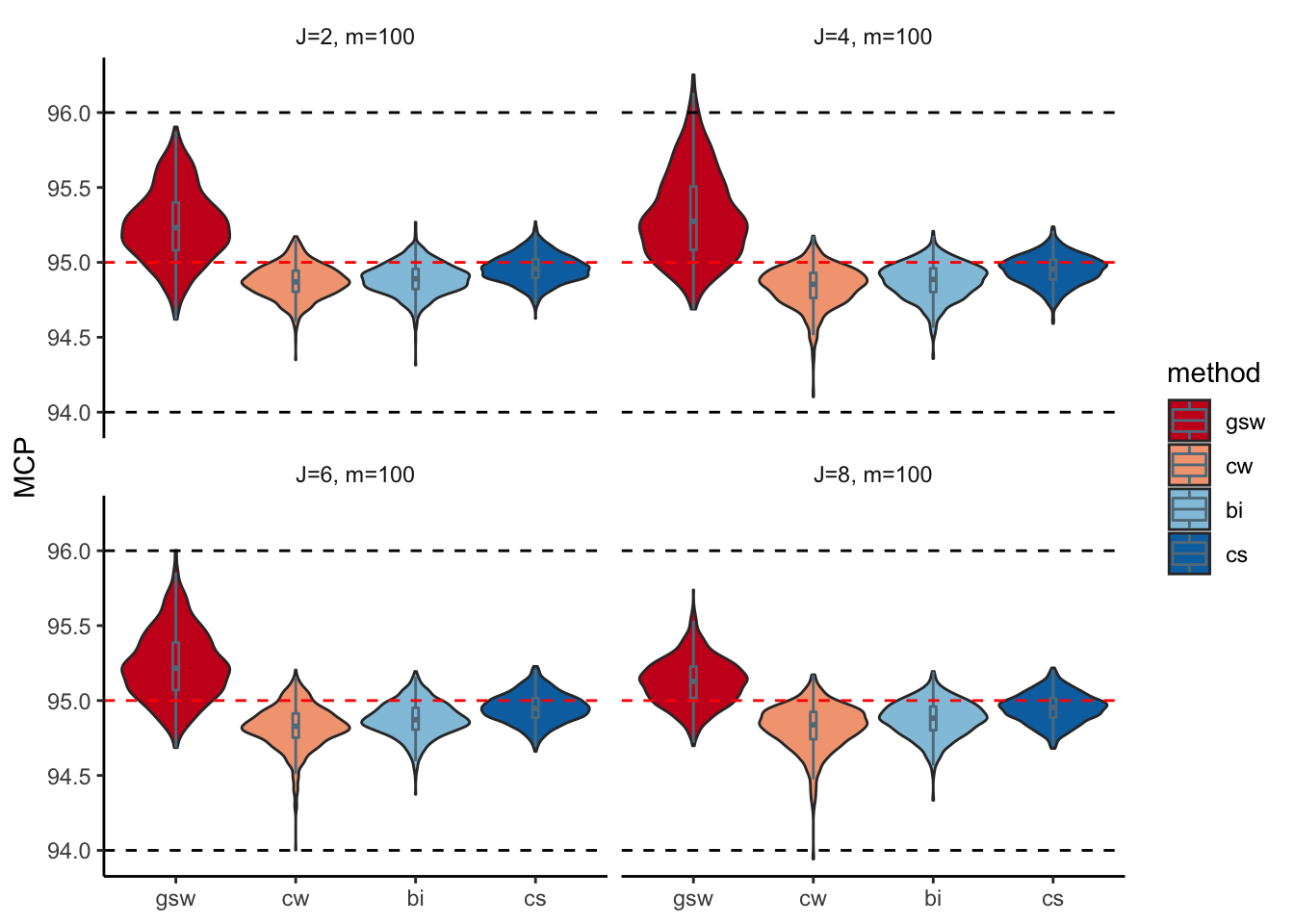}
    \caption{Violin-plots and Boxplots of MCP with m=100}
    \label{fig:galaxy}
\end{figure}

\begin{figure}[htp]
    \centering
    \includegraphics[width=16cm]{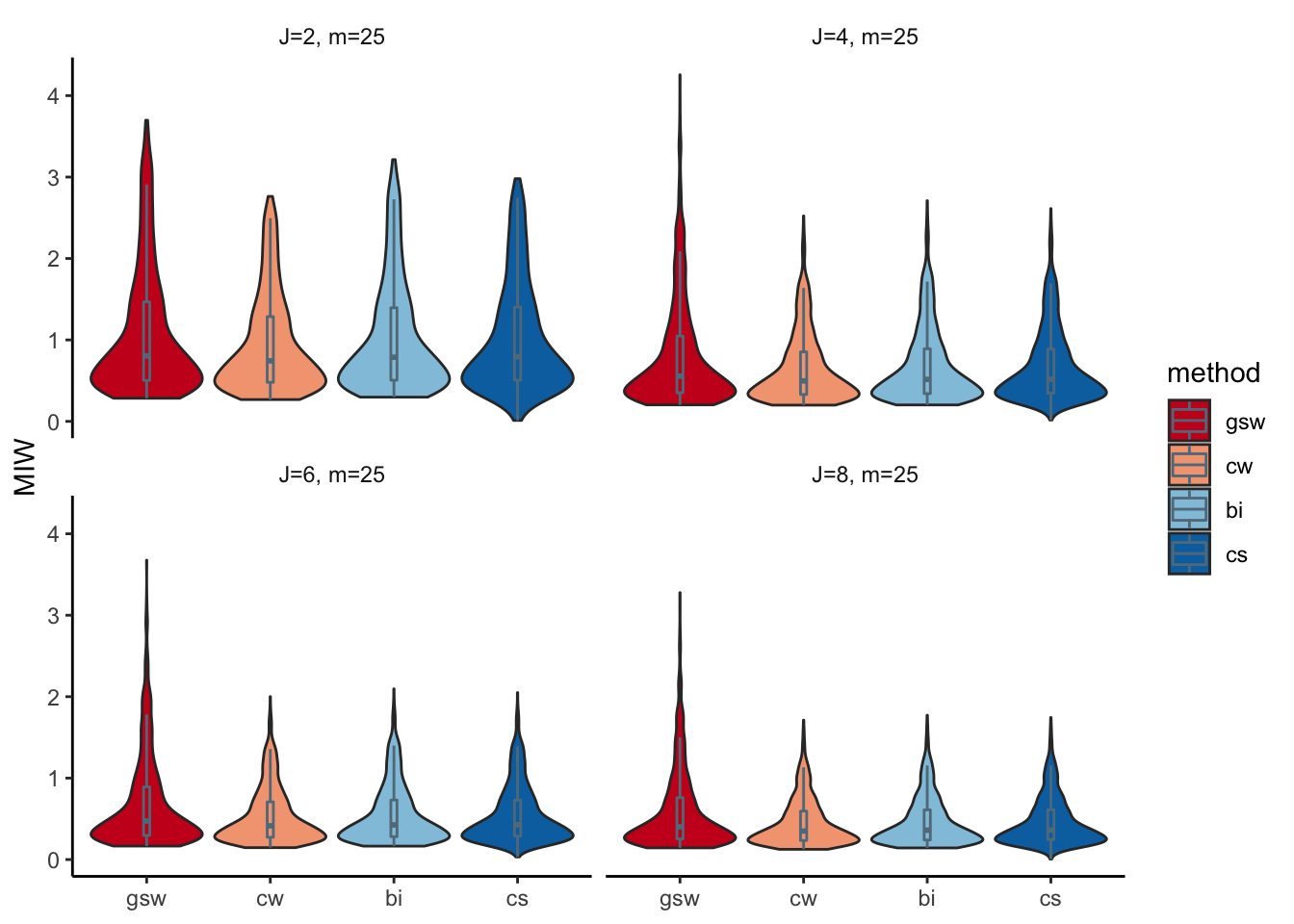}
    \caption{Violin-plots and Boxplots of MIW with m=25}
    \label{fig:galaxy}
\end{figure}

\begin{figure}[htp]
    \centering
    \includegraphics[width=16cm]{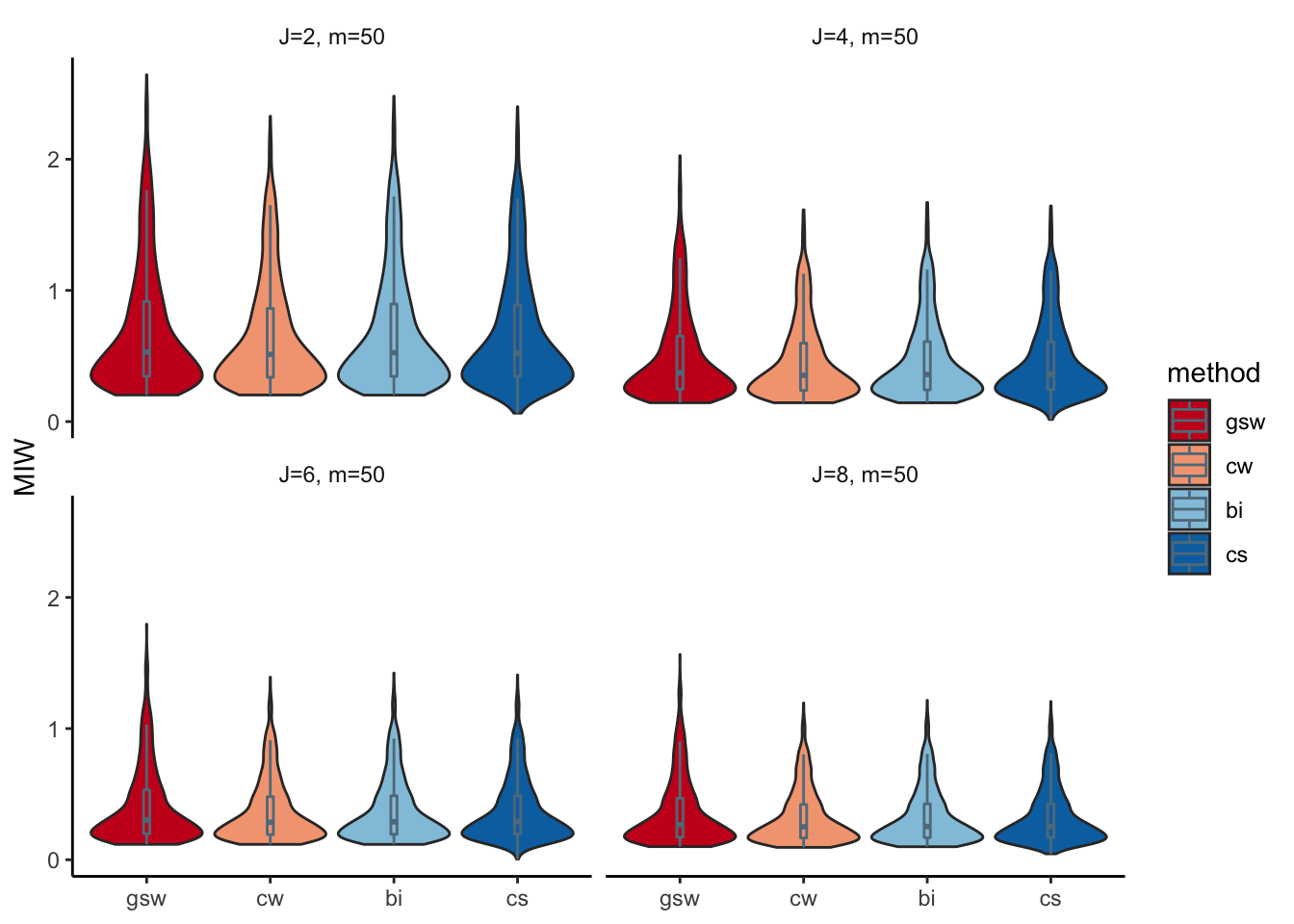}
    \caption{Violin-plots and Boxplots of MIW with m=50}
    \label{fig:galaxy}
\end{figure}

\begin{figure}[htp]
    \centering
    \includegraphics[width=16cm]{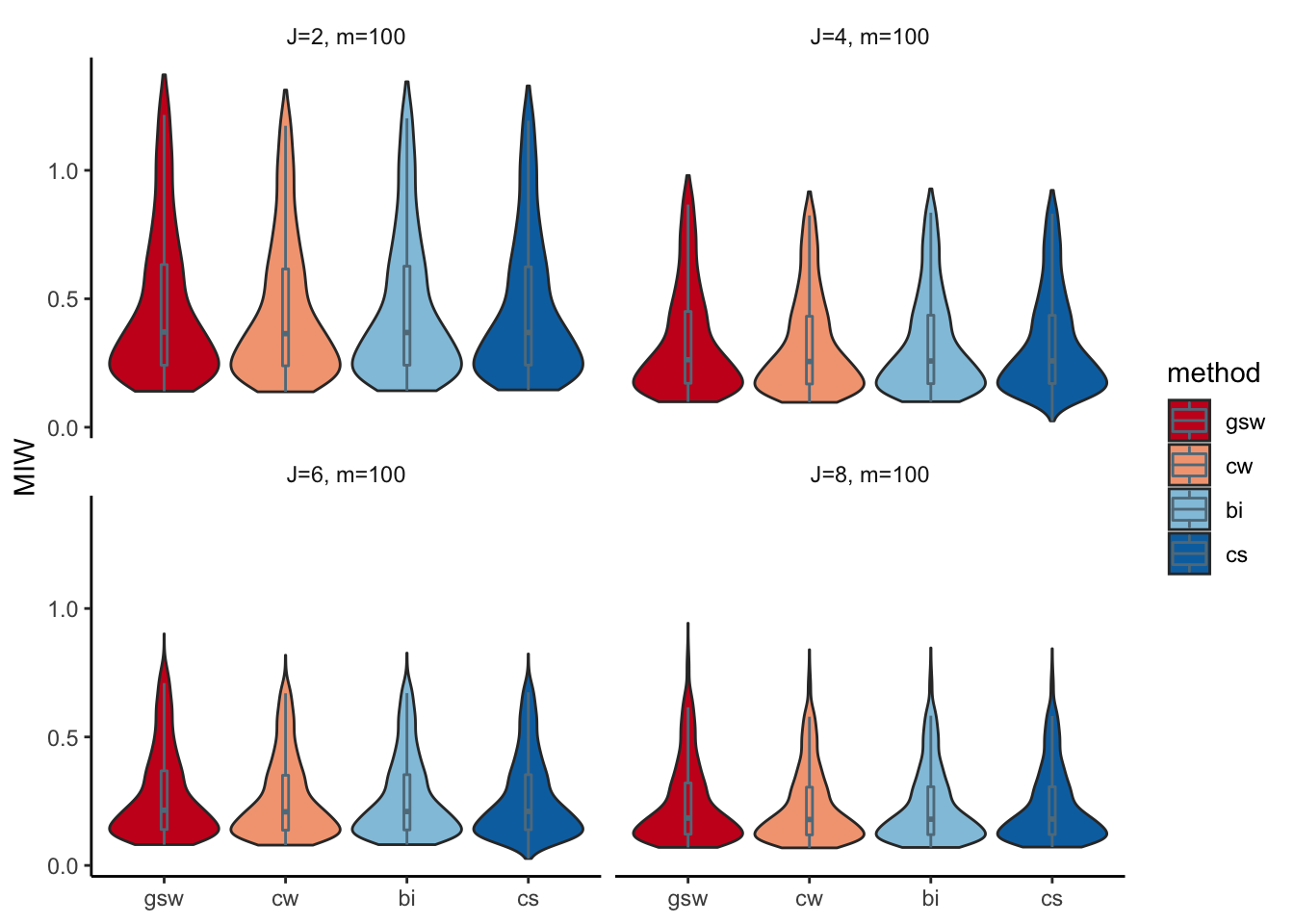}
    \caption{Violin-plots and Boxplots of MIW with m=100}
    \label{fig:galaxy}
\end{figure}

\begin{figure}[htp]
    \centering
    \includegraphics[width=16cm]{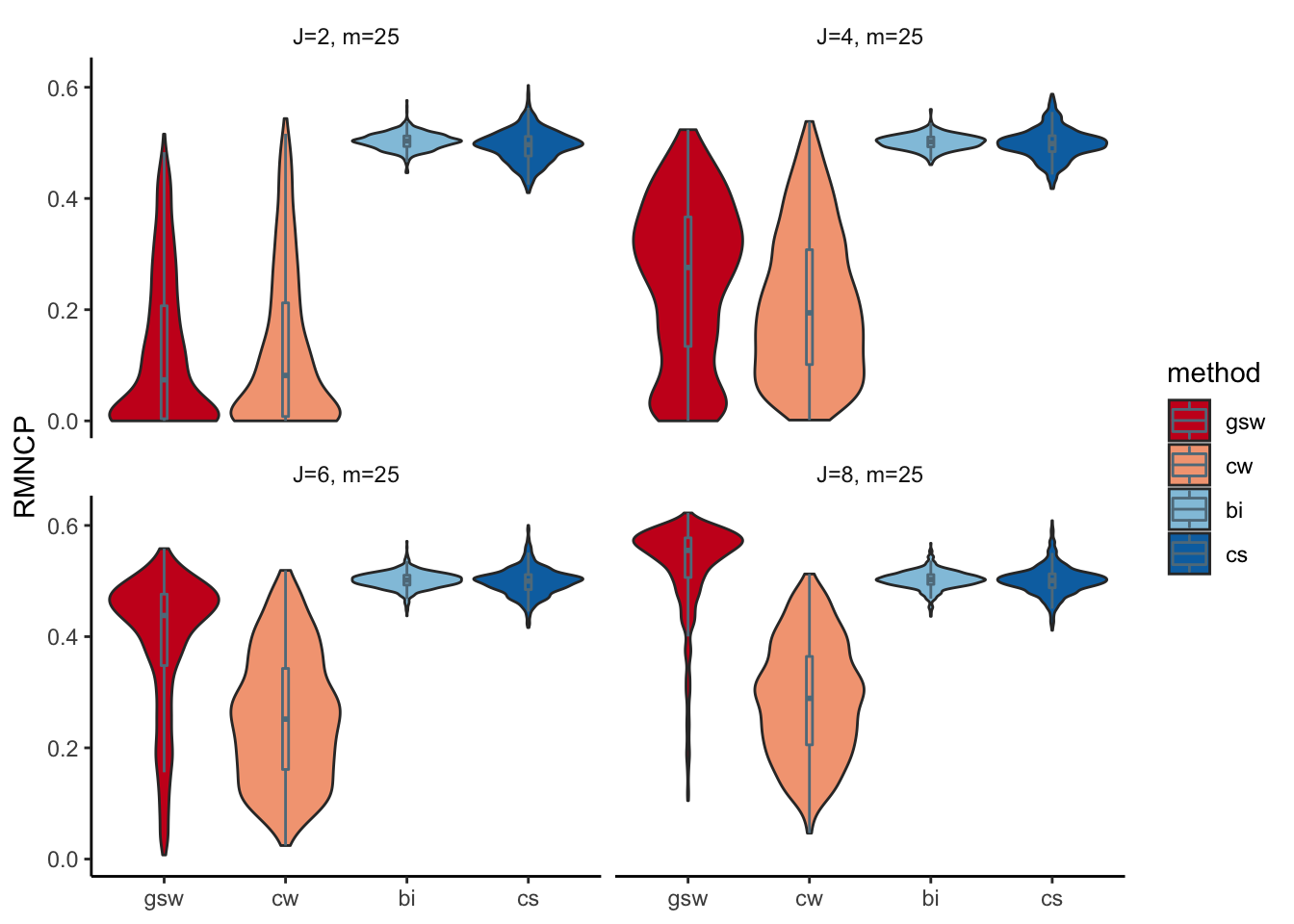}
    \caption{Violin-plots and Boxplots of RMNCP with m=25}
    \label{fig:galaxy}
\end{figure}

\begin{figure}[htp]
    \centering
    \includegraphics[width=16cm]{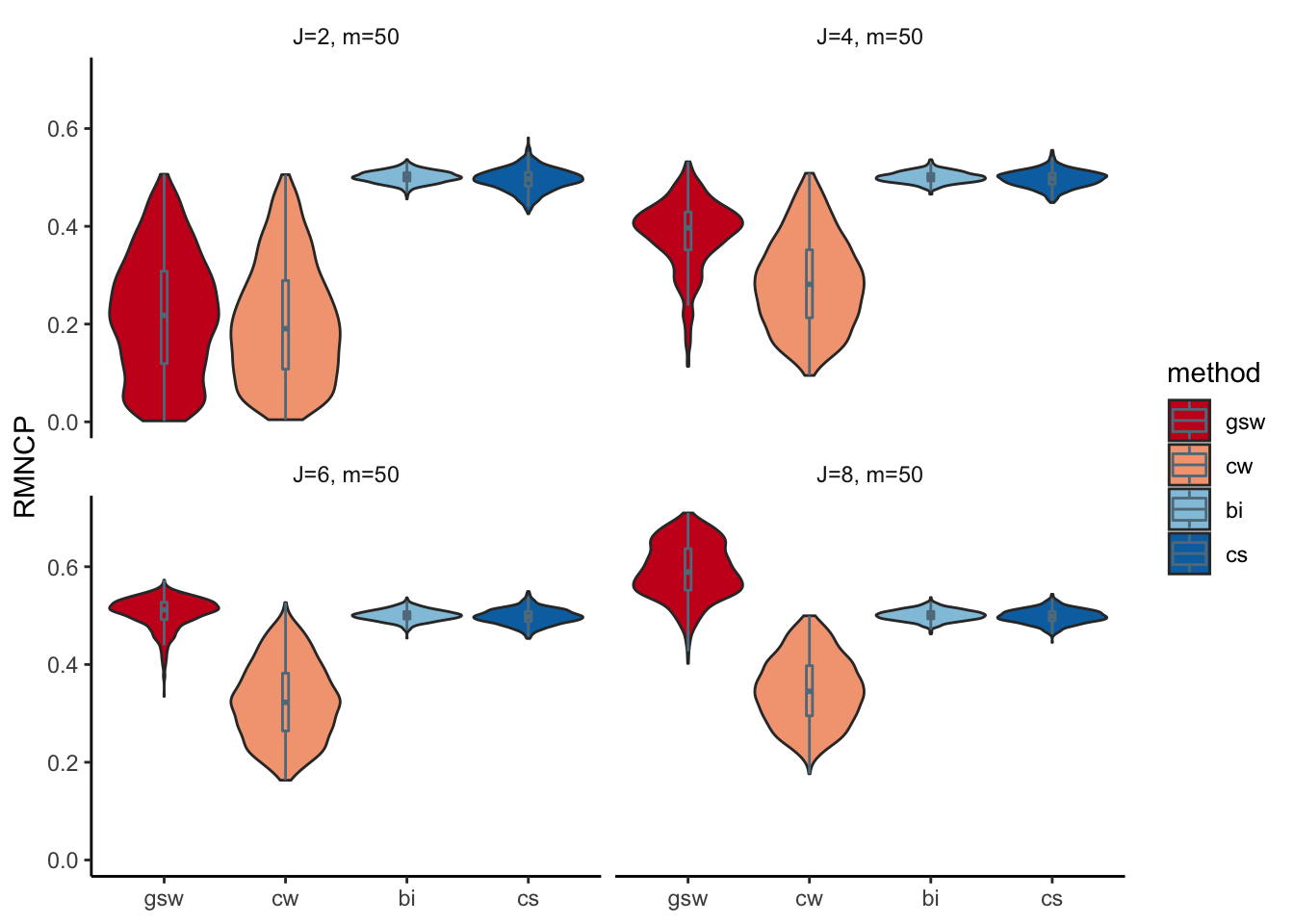}
    \caption{Violin-plots and Boxplots of RMNCP with m=50}
    \label{fig:galaxy}
\end{figure}

\begin{figure}[htp]
    \centering
    \includegraphics[width=16cm]{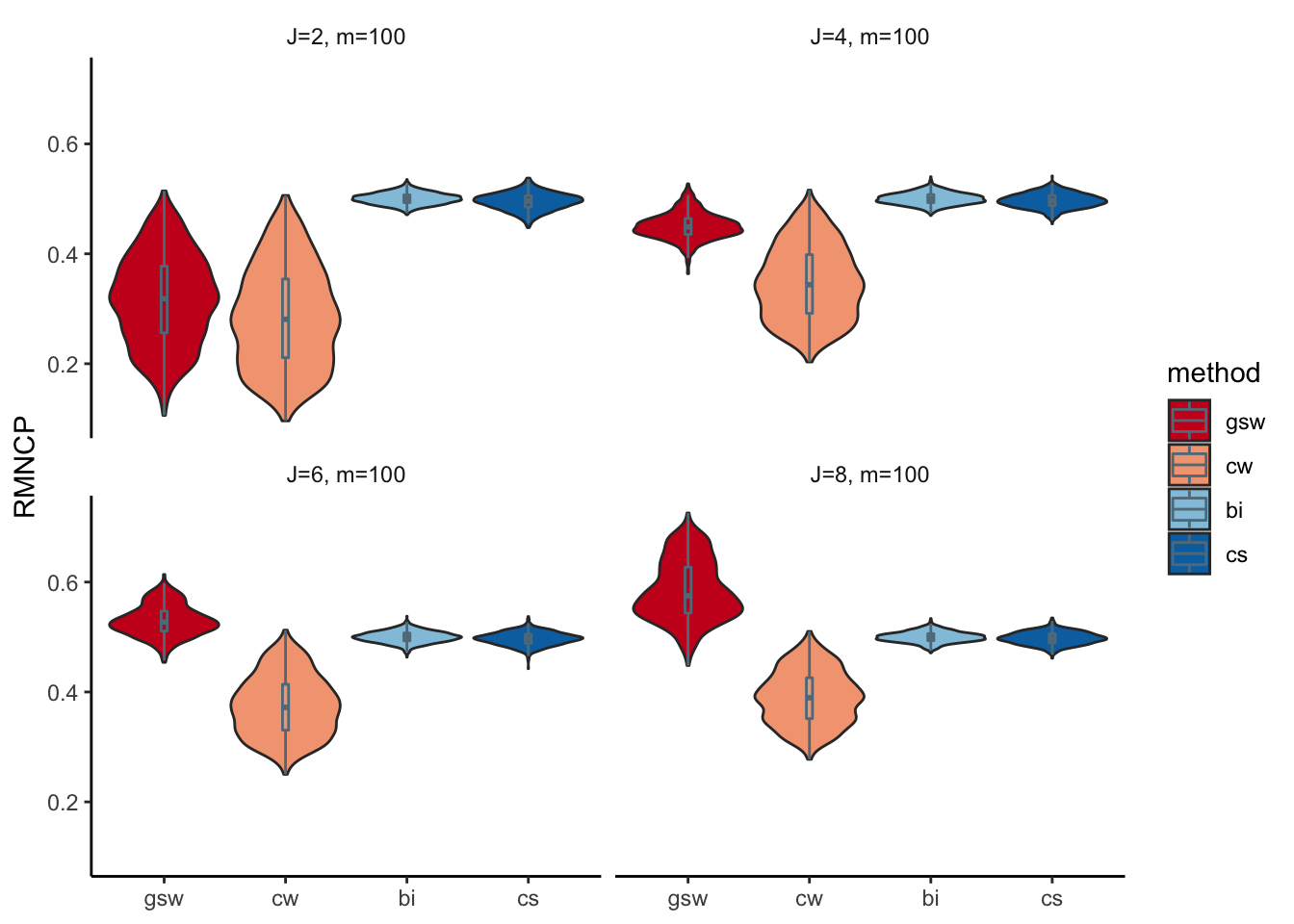}
    \caption{Violin-plots and Boxplots of RMNCP with m=100}
    \label{fig:galaxy}
\end{figure}

\section{Real Data Example}

We use the double-blinded randomized clinical trial data by Mandel et al. (1982) to illustrate the proposed four CI constructions. In this clinical trial, children who suffer from otitis media with effusion (OME) and have bilateral tympanocentesis are randomly assigned into two groups: amoxicillin or cefaclor, to receive the 14-day treatment ~\cite{mandel1982duration}. After the treatment, the number of cured ears are summarized in table 18 with age group as strata. 

By applying the four CI constructions in section 4, we obtain the MLEs of parameters based on observed data, shown in Table 19, and the four CI estimators are summarized in Table 20. As shown, all the CIs include 1, which means there is no statistical evidence to reject null hypothesis $H_0: \delta_1=\delta_2=\cdots=\delta_J=1$.

\begin{singlespace}
\begin{longtable}[t]{rrrrrrr}
\caption{\label{tab:unnamed-chunk-2} Frequency of number of OME-free ears after treatment.}\\
\toprule
\multicolumn{1}{c}{} & \multicolumn{6}{c}{Age group} \\
\cmidrule(l{3pt}r{3pt}){2-7}
\multicolumn{1}{c}{} & \multicolumn{2}{c}{ $<$2 years} & \multicolumn{2}{c}{2-5 years} & \multicolumn{2}{c}{$\geq$6 years} \\
\cmidrule(l{3pt}r{3pt}){2-3} \cmidrule(l{3pt}r{3pt}){4-5} \cmidrule(l{3pt}r{3pt}){6-7}
Number of OME-Free ears & Cefactor & Amoxicillin & Cefactor & Amoxicillin & Cefactor & Amoxicillin\\
\midrule
\endfirsthead
\caption[]{Table 1. Frequency of number of OME-free ears after treatment. \textit{(continued)}}\\
\toprule
\multicolumn{1}{c}{} & \multicolumn{6}{c}{Age group} \\
\cmidrule(l{3pt}r{3pt}){2-7}
\multicolumn{1}{c}{} & \multicolumn{2}{c}{<2 years} & \multicolumn{2}{c}{2-5 years} & \multicolumn{2}{c}{$\geq$6 years} \\
\cmidrule(l{3pt}r{3pt}){2-3} \cmidrule(l{3pt}r{3pt}){4-5} \cmidrule(l{3pt}r{3pt}){6-7}
Number of OME-Free ears & Cefactor & Amoxicillin & Cefactor & Amoxicillin & Cefactor & Amoxicillin\\
\midrule
\endhead

\endfoot
\bottomrule
\endlastfoot
0 & 8 & 11 & 6 & 3 & 0 & 1\\
1 & 2 & 2 & 6 & 1 & 1 & 0\\
2 & 8 & 2 & 10 & 5 & 3 & 6\\*
\end{longtable}
\end{singlespace}

\begin{singlespace}
\begin{longtable}[t]{lrrrrrlrrl}
\caption{\label{tab:unnamed-chunk-4}MlEs of parameters based on observed data.}\\
\toprule
\multicolumn{1}{c}{} & \multicolumn{3}{c}{Global MlEs} & \multicolumn{3}{c}{Unconstrained MLEs} & \multicolumn{3}{c}{Constrained MLEs} \\
\cmidrule(l{3pt}r{3pt}){2-4} \cmidrule(l{3pt}r{3pt}){5-7} \cmidrule(l{3pt}r{3pt}){8-10}
\multicolumn{1}{l}{Age groups} & \multicolumn{1}{l}{$\tilde{\pi}_{1}$} & \multicolumn{1}{l}{$\tilde{\gamma}$} & \multicolumn{1}{l}{$\tilde{\delta}$} & \multicolumn{1}{l}{$\hat{\pi}_{1}$} & \multicolumn{1}{l}{$\hat{\gamma}$} & \multicolumn{1}{l}{$\hat{\delta}$} & \multicolumn{1}{l}{$\hat{\pi}_{1{H_{0}}}$} & \multicolumn{1}{l}{$\hat{\gamma}_{H_{0}}$} & \multicolumn{1}{l}{$\hat{\delta}_{H_{0}}$}\\
\midrule
\endfirsthead
\caption[]{MlEs of parameters based on observed data. \textit{(continued)}}\\
\toprule
\multicolumn{1}{c}{} & \multicolumn{3}{c}{Global MlEs} & \multicolumn{3}{c}{Unconstrained MLEs} & \multicolumn{3}{c}{Constrained MLEs} \\
\cmidrule(l{3pt}r{3pt}){2-4} \cmidrule(l{3pt}r{3pt}){5-7} \cmidrule(l{3pt}r{3pt}){8-10}
\multicolumn{1}{l}{Age groups} & \multicolumn{1}{l}{$\tilde{\pi}_{1}$} & \multicolumn{1}{l}{$\tilde{\gamma}$} & \multicolumn{1}{l}{$\tilde{\delta}$} & \multicolumn{1}{l}{$\hat{\pi}_{1}$} & \multicolumn{1}{l}{$\hat{\gamma}$} & \multicolumn{1}{l}{$\hat{\delta}$} & \multicolumn{1}{l}{$\hat{\pi}_{1{H_{0}}}$} & \multicolumn{1}{l}{$\hat{\gamma}_{H_{0}}$} & \multicolumn{1}{l}{$\hat{\delta}_{H_{0}}$}\\
\midrule
\endhead

\endfoot
\bottomrule
\endlastfoot
\multicolumn{1}{l}{$<$2 years} & \multicolumn{1}{l}{0.476} & \multicolumn{1}{l}{0.833} & \multicolumn{1}{l}{0.480} & \multicolumn{1}{l}{0.404} & \multicolumn{1}{l}{0.833} & \multicolumn{1}{l}{0.8174} & \multicolumn{1}{l}{0.364} & \multicolumn{1}{l}{0.833} & \multicolumn{1}{l}{1}\\
\multicolumn{1}{l}{2-5 years} & \multicolumn{1}{l}{0.612} & \multicolumn{1}{l}{0.811} & \multicolumn{1}{l}{0.917} & \multicolumn{1}{l}{0.625} & \multicolumn{1}{l}{0.811} & \multicolumn{1}{l}{-} & \multicolumn{1}{l}{0.597} & \multicolumn{1}{l}{0.811} & \multicolumn{1}{l}{-}\\
\multicolumn{1}{l}{$\geq$6 years} & \multicolumn{1}{l}{0.950} & \multicolumn{1}{l}{0.947} & \multicolumn{1}{l}{0.857} & \multicolumn{1}{l}{0.950} & \multicolumn{1}{l}{0.947} & \multicolumn{1}{l}{-} & \multicolumn{1}{l}{0.864} & \multicolumn{1}{l}{0.947} & \multicolumn{1}{l}{-}\\*
\end{longtable}
\end{singlespace}

\begin{singlespace}
\begin{longtable}[t]{lrr}
\caption{\label{tab:unnamed-chunk-5}Confidence interval estimation of proportion ratios for the proposed methods.}\\
\toprule
method & Lower Limit & Upper Limit\\
\midrule
\endfirsthead
\caption[]{Confidence interval estimation of proportion ratios for the proposed methods. \textit{(continued)}}\\
\toprule
method & Lower Limit & Upper Limit\\
\midrule
\endhead

\endfoot
\bottomrule
\endlastfoot
Global Weighted Wald (SGW) & 0.432 & 1.000\\
Profile Likelihood (BI) & 0.529 & 1.047\\
Complete Wald (CW) & 0.548 & 1.080\\
Complete Score (CS) & 0.529 & 1.113\\*
\end{longtable}
\end{singlespace}

\section{Conclusions}

Following Dallal's intraclass correlation model, this article utilizes four confidence interval methods (SGW, CW, BI, CS) to solve the CI construction for proportion ratios of two proportions on stratified bilateral correlated data. In addition, three evaluations indices (MCP, MIW, RMNCP) are proposed to generate the relative comprehensive evaluation for four presented confidence interval methods. 
  
 Two Monte Carlo simulation results show that profile likelihood confidence interval with the bisection algorithm (BI) is recommended without considering the conservative performance in small sample size with multiple strata. Meanwhile, the complete MLE-based score confidence interval (CS) method outperforms other CI methods considering all three evaluation indices, which is highly recommended. 
 
However, asymptotic methods might be limited because of poor performance under a small sample size. Future work might consider exact tests to investigate related issues.

\section{Appendix}
\subsection{Information matrix derivation for sample-size weighted global-MLE based wald-type confidence interval (SGW)}

The first order differential equations of stratum j are:

$$\frac{\partial l_{j}}{\partial \delta_{j}}=
\frac{ m_{12j} }{ \delta_{j}}+\frac{ m_{22j} }{ \delta_{j}}+\frac{ m_{02j} \, \pi_{1 j } \,{\left(\gamma_j -2\right)}}{ \delta_{j}\, \pi_{1 j } \,{\left(\gamma_j -2\right)}+1} , $$

$$\frac{\partial l_{j}}{\partial \pi_{1 j}} = \frac{ m_{12j} }{ \pi_{1 j } }+\frac{ m_{21j} }{ \pi_{1 j } }+\frac{ m_{22j} }{ \pi_{1 j } }+\frac{ m_{01j} \,{\left(\gamma_j -2\right)}}{ \pi_{1 j } \,{\left(\gamma_j -2\right)}+1}+\frac{ \delta_{j}\, m_{02j} \,{\left(\gamma_j -2\right)}}{ \delta_{j}\, \pi_{1 j } \,{\left(\gamma_j -2\right)}+1}+\frac{ m_{11j} \,{\left(2\,\gamma_j -2\right)}}{2\, \pi_{1 j } \,{\left(\gamma_j -1\right)}} , $$

$$\frac{\partial l_{j}}{\partial \gamma_{j}} = \frac{ m_{21j} }{\gamma_j }+\frac{ m_{22j} }{\gamma_j }+\frac{ m_{11j} }{\gamma_j -1}+\frac{ m_{12j} }{\gamma_j -1}+\frac{ m_{01j} \, \pi_{1 j } }{ \pi_{1 j } \,{\left(\gamma_j -2\right)}+1}+\frac{ \delta_{j}\, m_{02j} \, \pi_{1 j } }{ \delta_{j}\, \pi_{1 j } \,{\left(\gamma_j -2\right)}+1} . $$

The second order differential equations of stratum j are:

$$ \frac{\partial^{2} l_{j}}{\partial \delta_{j}^{2}} = 
-\frac{ m_{12j} }{ \delta_{j}^2 }-\frac{ m_{22j} }{ \delta_{j}^2 }-\frac{ m_{02j} \,{ \pi_{1 j } }^2 \,{{\left(\gamma_j -2\right)}}^2 }{{{\left( \delta_{j} \, \pi_{1 j } \,{\left(\gamma_j -2\right)}+1\right)}}^2 } , $$

$$\frac{\partial^{2} l_{j}}{\partial \delta_{j} \partial \pi_{1 j}}=
\frac{ m_{02j} \,{\left(\gamma_j -2\right)}}{ \delta_{j}\, \pi_{1 j } \,{\left(\gamma_j -2\right)}+1}-\frac{ \delta_{j}\, m_{02j} \, \pi_{1 j } \,{{\left(\gamma_j -2\right)}}^2 }{{{\left( \delta_{j}\, \pi_{1 j } \,{\left(\gamma_j -2\right)}+1\right)}}^2 } , $$

$$\frac{\partial^{2} l_{j}}{\partial \delta_{j} \partial \gamma_{j}}= \frac{ m_{02j} \, \pi_{1 j } }{ \delta_{j}\, \pi_{1 j } \,{\left(\gamma_j -2\right)}+1}-\frac{ \delta_{j}\, m_{02j} \,{ \pi_{1 j } }^2 \,{\left(\gamma_j -2\right)}}{{{\left( \delta_{j}\, \pi_{1 j } \,{\left(\gamma_j -2\right)}+1\right)}}^2 } , $$

$$\frac{\partial^{2} l_{j}}{ \partial \pi_{1 j} \partial \delta_{j}} =\frac{ m_{02j} \,{\left(\gamma_j -2\right)}}{ \delta_{j}\, \pi_{1 j } \,{\left(\gamma_j -2\right)}+1}-\frac{ \delta_{j}\, m_{02j} \, \pi_{1 j } \,{{\left(\gamma_j -2\right)}}^2 }{{{\left( \delta_{j}\, \pi_{1 j } \,{\left(\gamma_j -2\right)}+1\right)}}^2 } , $$

$$\frac{\partial^{2} l_{j}}{\partial \pi_{1 j}^{2}} = -\frac{ m_{12j} }{{ \pi_{1 j } }^2 }-\frac{ m_{21j} }{{ \pi_{1 j } }^2 }-\frac{ m_{22j} }{{ \pi_{1 j } }^2 }-\frac{ m_{01j} \,{{\left(\gamma_j -2\right)}}^2 }{{{\left( \pi_{1 j } \,{\left(\gamma_j -2\right)}+1\right)}}^2 }-\frac{\delta^2 \, m_{02j} \,{{\left(\gamma_j -2\right)}}^2 }{{{\left( \delta_{j}\, \pi_{1 j } \,{\left(\gamma_j -2\right)}+1\right)}}^2 }-\frac{ m_{11j} \,{\left(2\,\gamma_j -2\right)}}{2\,{ \pi_{1 j } }^2 \,{\left(\gamma_j -1\right)}} , $$

$$\frac{\partial^{2} l_{j}}{\partial \pi_{1 j} \partial \gamma_{j}} = \frac{ m_{01j} }{{{\left( \pi_{1 j } \,{\left(\gamma_j -2\right)}+1\right)}}^2 }+\frac{ \delta_{j}\, m_{02j} }{{{\left( \delta_{j}\, \pi_{1 j } \,{\left(\gamma_j -2\right)}+1\right)}}^2 } , $$

$$\frac{\partial^{2} l_{j}}{\partial \gamma_{j} \partial \delta_{j} } =\frac{ m_{02j} \, \pi_{1 j } }{ \delta_{j}\, \pi_{1 j } \,{\left(\gamma_j -2\right)}+1}-\frac{ \delta_{j}\, m_{02j} \,{ \pi_{1 j } }^2 \,{\left(\gamma_j -2\right)}}{{{\left( \delta_{j}\, \pi_{1 j } \,{\left(\gamma_j -2\right)}+1\right)}}^2 } , $$

$$\frac{\partial^{2} l_{j}}{\partial \gamma_{j} \partial \pi_{1 j} }
\frac{ m_{01j} }{{{\left( \pi_{1 j } \,{\left(\gamma_j -2\right)}+1\right)}}^2 }+\frac{ \delta_{j}\, m_{02j} }{{{\left( \pi_{1 j } \,{\left(2\, \delta_{j}- \delta_{j}\,\gamma_j \right)}-1\right)}}^2 } , $$

$$\frac{\partial^{2} l_{j}}{\partial \gamma_{ j}^{2}} =-\frac{ m_{21j} }{{\gamma_j }^2 }-\frac{ m_{22j} }{{\gamma_j }^2 }-\frac{ m_{11j} }{{{\left(\gamma_j -1\right)}}^2 }-\frac{ m_{12j} }{{{\left(\gamma_j -1\right)}}^2 }-\frac{ m_{01j} \,{ \pi_{1 j } }^2 }{{{\left( \pi_{1 j } \,{\left(\gamma_j -2\right)}+1\right)}}^2 }-\frac{\delta^2 \, m_{02j} \,{ \pi_{1 j } }^2 }{{{\left( \delta_{j}\, \pi_{1 j } \,{\left(\gamma_j -2\right)}+1\right)}}^2 } . $$








Fisher information of stratum j are: 

$$I_{11}^{(j)}=E\left(-\frac{\partial^{2} l_{j}}{\partial \delta_{j}^{2}}\right) = \frac{m_{+2j} }{ \delta_{j}^2 \,{\left( \delta_{j} \, \pi_{1 j } \,{\left(\gamma_j -2\right)}+1\right)}}-\frac{m_{+2j} }{ \delta_{j}^2 } , $$

$$I_{12}^{(j)}=I_{21}^{(j)}=E\left(-\frac{\partial^{2} l_{j}}{\partial \delta_{j} \partial \pi_{1 j}}\right)= \-\frac{m_{+2j} \,{\left(\gamma_j -2\right)}}{ \delta_{j} \, \pi_{1 j } \,{\left(\gamma_j -2\right)}+1} , $$

$$I_{13}^{(j)}=I_{31}^{(j)}=E\left(-\frac{\partial^{2} l_{j}}{\partial \delta_{j} \partial  \gamma_{j}}\right) = -\frac{m_{+2j} \, \pi_{1 j } }{ \delta_{j} \, \pi_{1 j } \,{\left(\gamma_j -2\right)}+1} , $$

$$I_{22}^{(j)}=E\left(-\frac{\partial^{2} l_{j}}{\partial \pi_{1 j}^{2}}\right)= \frac{m_{+1j} \,{{\left(\gamma_j -2\right)}}^2 }{ \pi_{1 j } \,{\left(\gamma_j -2\right)}+1}-\frac{{\left(\gamma_j -2\right)}\,{\left(m_{+1j} + \delta_{j} \,m_{+2j} \right)}}{ \pi_{1 j } }-\frac{ \delta_{j}^2 \,m_{+2j} \,{{\left(\gamma_j -2\right)}}^2 }{ \pi_{1 j } \,{\left(2\, \delta_{j} - \delta_{j} \,\gamma_j \right)}-1} , $$

$$I_{23}^{(j)}=I_{32}^{(j)}=E\left(-\frac{\partial^{2} l_{j}}{\partial \pi_{1 j} \gamma_{j}}\right) = \frac{ \delta_{j} \,m_{+2j} }{ \pi_{1 j } \,{\left(2\, \delta_{j} - \delta_{j} \,\gamma_j \right)}-1}-\frac{m_{+1j} }{ \pi_{1 j } \,{\left(\gamma_j -2\right)}+1} , $$

$$
\begin{aligned}
I_{33}^{(j)}=E\left(-\frac{\partial^{2} l_{j}}{\partial \gamma_{j}^{2}}\right) 
& = \frac{m_{+1j} \, \pi_{1 j } }{\gamma_j }
-\frac{2\,m_{+1j} \, \pi_{1 j } }{\gamma_j -1}
+\frac{m_{+1j} \,{ \pi_{1 j } }^2 }{ \pi_{1 j } \,{\left(\gamma_j -2\right)}+1} \\
& +\frac{ \delta_{j}^2 \,m_{+2j} \,{ \pi_{1 j } }^2 }{ \delta_{j} \, \pi_{1 j } \,{\left(\gamma_j -2\right)}+1}
+\frac{ \delta_{j} \,m_{+2j} \, \pi_{1 j } }{\gamma_j }
-\frac{2\, \delta_{j} \,m_{+2j} \, \pi_{1 j } }{\gamma_j -1}. 
\end{aligned}
$$

and $$I_j=\left[\begin{array}{ccc}I_{ 11}^{(j)} & I_{ 12}^{(j)} & I_{ 13}^{(j)} \\ I_{ 21}^{(j)} & I_{ 22}^{(j)} & I_{ 23}^{(j)} \\ I_{ 31}^{(j)} & I_{ 32}^{(j)} & I_{ 33}^{(j)}\end{array}\right]_{3J \times 3J}.$$

Then

$$I_j^{-1} = \left (\frac{1} {D ^{(j)}} \right) \left(\begin{array}{ccc}
{\left(I_{22 }^{(j)} \,I_{33 }^{(j)} -I_{23 }^{(j)} \,I_{32 }^{(j)} \right) }\, & -{\left(I_{12 }^{(j)} \,I_{33 }^{(j)} -I_{13 }^{(j)} \,I_{32 }^{(j)} \right) }\, & {\left(I_{12 }^{(j)} \,I_{23 }^{(j)} -I_{13 }^{(j)} \,I_{22 }^{(j)} \right) }\,  \\
-{\left(I_{21 }^{(j)} \,I_{33 }^{(j)} -I_{23 }^{(j)} \,I_{31 }^{(j)} \right) }\,  & {\left(I_{11 }^{(j)} \,I_{33 }^{(j)} -I_{13 }^{(j)} \,I_{31 }^{(j)} \right) }\,   & -{\left(I_{11 }^{(j)} \,I_{23 }^{(j)} -I_{13 }^{(j)} \,I_{21 }^{(j)} \right) }\, \\
{\left(I_{21 }^{(j)} \,I_{32 }^{(j)} -I_{22 }^{(j)} \,I_{31 }^{(j)} \right) }\, & -{\left(I_{11 }^{(j)} \,I_{32 }^{(j)} -I_{12 }^{(j)} \,I_{31 }^{(j)} \right) }\,   & {\left(I_{11 }^{(j)} \,I_{22 }^{(j)} -I_{12 }^{(j)} \,I_{21 }^{(j)} \right) }\,      
\end{array}\right) , $$ and 

$$D ^{(j)}= I_{11 }^{(j)} \,I_{22 }^{(j)} \,I_{33 }^{(j)} -I_{11 }^{(j)} \,I_{23 }^{(j)} \,I_{32 }^{(j)} -I_{12 }^{(j)} \,I_{21 }^{(j)} \,I_{33 }^{(j)} +I_{12 }^{(j)} \,I_{23 }^{(j)} \,I_{31 }^{(j)} +I_{13 }^{(j)} \,I_{21 }^{(j)} \,I_{32 }^{(j)} -I_{13 }^{(j)} \,I_{22 }^{(j)} \,I_{31 }^{(j)} .$$

Then the $c I_j^{-1} c^T=I_j^{-1}{ }_{(1,1)}=\left(\frac{1}{D^{(j)}}\right)\left(I_{22}^{(j)} I_{33}^{(j)}-I_{23}^{(j)} I_{32}^{(j)}\right)$

\subsection{Information matrix derivation for complete MLE-based Wald-type confidence interval and Score confidence interval}

Under the assumption that each stratum has a common ratio of proportions $\delta$, the Fisher information matrix is written as

$$I=\left[
\begin{array}{ccc}
\underset{(1 \times 1)}{I_{11}} & \underset{(1 \times J)}{I_{12}} & \underset{(1 \times J)}{I_{13}} \\ 
\underset{(J \times 1)}{I_{21}} & \underset{(J \times J)}{I_{22}} & \underset{(J \times J)}{I_{23}} \\ 
\underset{(J \times 1)}{I_{31}} & \underset{(J \times J)}{I_{32}} & \underset{(J \times J)}{I_{33}}
\end{array}
\right]$$

where 

$$
I_{11}=\mathrm{E}\left(-\sum_{j=1}^J \frac{\partial^2 l_j}{\partial \delta^2}\right)= \sum_{j=1}^J \frac{m_{+2j} }{ \delta^2 \,{\left( \delta \, \pi_{1 j } \,{\left(\gamma_j -2\right)}+1\right)}}-\frac{m_{+2j} }{ \delta^2 } , 
$$

$$
\begin{aligned}
I_{12} = I_{21}^T 
& = \left[\mathrm{E}\left(-\frac{\partial^2 l_1}{\partial \delta \partial \pi_{11}}\right), \mathrm{E}\left(-\frac{\partial^2 l_2}{\partial \delta \partial \pi_{12}}\right), \ldots, \mathrm{E}\left(-\frac{\partial^2 l_J}{\partial \delta \partial \pi_{1 J}}\right)\right]\\
&=\left[I_{12}^{(1)}, I_{12}^{(2)}, \ldots, I_{12}^{(J)}\right]
\end{aligned}
$$

Where 
$$
I_{12 }^{(j)} = \-\frac{m_{+2j} \,{\left(\gamma_j -2\right)}}{ \delta \, \pi_{1 j } \,{\left(\gamma_j -2\right)}+1} ,
$$

$$
\begin{aligned}
I_{13} = I_{31}^T 
& = \left[\mathrm{E}\left(-\frac{\partial^2 l_1}{\partial \delta \partial \gamma_{1}}\right), \mathrm{E}\left(-\frac{\partial^2 l_2}{\partial \delta \partial \gamma_{2}}\right), \ldots, \mathrm{E}\left(-\frac{\partial^2 l_J}{\partial \delta \partial \gamma_{ J}}\right)\right]\\
&=\left[I_{13}^{(1)}, I_{13}^{(2)}, \ldots, I_{13}^{(J)}\right]
\end{aligned}
$$

where

$$I_{13}^{(j)}= -\frac{m_{+2j} \, \pi_{1 j } }{ \delta \, \pi_{1 j } \,{\left(\gamma_j -2\right)}+1}. $$

$$
I_{22}=\operatorname{diag}\left\{\mathrm{E}\left(-\frac{\partial^2 l_j}{\partial \pi_{1 j}^2}\right)\right\}=\left(\begin{array}{llll}
I_{22}^{(1)} & & & \\
& I_{22}^{(2)} & & \\
& & \ddots & \\
& & & I_{22}^{(J)}
\end{array}\right), \quad j=1,2, \ldots, J
$$

where

$$I_{22}^{(j)}=\frac{m_{+1j} \,{{\left(\gamma_j -2\right)}}^2 }{ \pi_{1 j } \,{\left(\gamma_j -2\right)}+1}-\frac{{\left(\gamma_j -2\right)}\,{\left(m_{+1j} + \delta \,m_{+2j} \right)}}{ \pi_{1 j } }-\frac{ \delta^2 \,m_{+2j} \,{{\left(\gamma_j -2\right)}}^2 }{ \pi_{1 j } \,{\left(2\, \delta - \delta \,\gamma_j \right)}-1} , $$

$$
I_{32}^T=I_{23}=\operatorname{diag}\left\{E\left(-\frac{\partial^2 I_j}{\partial \pi_{1 j} \gamma_j}\right)\right\}=\left(\begin{array}{llll}
I_{23}^{(1)} & & & \\
& I_{23}^{(2)} & & \\
& & \ddots & \\
& & & I_{23}^{(J)}
\end{array}\right), \quad j=1,2, \ldots, J
$$

$$I_{23}^{(j)}=\frac{ \delta \,m_{+2j} }{ \pi_{1 j } \,{\left(2\, \delta - \delta \,\gamma_j \right)}-1}-\frac{m_{+1j} }{ \pi_{1 j } \,{\left(\gamma_j -2\right)}+1} , $$

$$
I_{33}=\operatorname{diag}\left\{\mathrm{E}\left(-\frac{\partial^2 l_j}{\partial \gamma_j^2}\right)\right\}=\left(\begin{array}{llll}
I_{33}^{(1)} & & & \\
& I_{33}^{(2)} & & \\
& & \ddots & \\
& & & I_{33}^{(J)}
\end{array}\right), \quad j=1,2, \ldots, J
$$

where 

$$
\begin{aligned}
I_{33}^{(j)}
& = \frac{m_{+1j} \, \pi_{1 j } }{\gamma_j }
-\frac{2\,m_{+1j} \, \pi_{1 j } }{\gamma_j -1}
+\frac{m_{+1j} \,{ \pi_{1 j } }^2 }{ \pi_{1 j } \,{\left(\gamma_j -2\right)}+1} \\
& +\frac{ \delta^2 \,m_{+2j} \,{ \pi_{1 j } }^2 }{ \delta \, \pi_{1 j } \,{\left(\gamma_j -2\right)}+1}+\frac{ \delta \,m_{+2j} \, \pi_{1 j } }{\gamma_j }-\frac{2\, \delta \,m_{+2j} \, \pi_{1 j } }{\gamma_j -1} . 
\end{aligned}
$$

Then

$$C I^{-1} C^T = I^{-1}_{(1,1)} = \left(I_{11}-\left[\begin{array}{ll}
I_{12} & I_{13}
\end{array}\right]\left[\begin{array}{ll}
I_{22} & I_{23} \\
I_{32} & I_{33}
\end{array}\right]^{-1}
\left[\begin{array}{l}
I_{21} \\
I_{31}
\end{array}\right]\right)^{-1}$$

\bibliographystyle{unsrt}
\bibliography{1.bib}
\end{document}